\documentclass[preprintnumbers,article,amsmath,amssymb,floatfix,10pt,prd,superscriptaddress,nofootinbib,showkeys]{revtex4}

\usepackage{doi}
\usepackage{hyperref}
\hypersetup{
	colorlinks=true,        
	linkcolor=blue,         
	citecolor=cyan,         
}

\usepackage{graphicx}
\usepackage{dcolumn}
\usepackage{bm}
\usepackage{color}
\usepackage{enumitem}
\usepackage{amsmath}
\usepackage{amssymb}

\newcommand{\beq}{\begin{equation}}
\newcommand{\eeq}{\end{equation}}
\newcommand{\bea}{\begin{eqnarray}}
\newcommand{\eea}{\end{eqnarray}}

\usepackage{bbm}
\usepackage{amsfonts}
\usepackage{mathrsfs}
\usepackage{latexsym}
\usepackage{epsfig}
\usepackage{epstopdf}
\usepackage{epstopdf}
\usepackage{graphicx}
\usepackage{amssymb}
\usepackage{amsmath}
\usepackage{dcolumn}
\usepackage{bm}
\usepackage{color}
\usepackage{comment}
\usepackage{float}
\usepackage{xcolor}
\usepackage{orcidlink}
\begin{document}

\preprint{}

\title{Deflection of light by dark matter supporting traversable wormholes in the framework of Kalb-Ramond gravity}

\author{Nayan Sarkar\orcidlink{0000-0002-3489-6509}}
\email{ nayan.mathju@gmail.com}
\affiliation{Department of Mathematics, Karimpur Pannadevi College, Karimpur-741152, Nadia, West Bengal, India}

\author{Susmita Sarkar\orcidlink{0009-0007-1179-2495}}
\email{ susmita.mathju@gmail.com}
\affiliation{Department of Applied Science and Humanities, Haldia Institute of Technology, Haldia-721657, Purba Medinipur, West Bengal, India}

\begin{abstract}
This study explores the strong deflection of light by the asymptotically flat traversable wormholes within dark matter halos under the framework of Kalb-Ramond gravity. In this context, first, we derive traversable wormhole solutions based on the King and Navarro-Frenk-White dark matter density profiles associated with anisotropic matter sources. For a particular set of parameters, the proposed shape functions are found to be positively increasing and satisfy all the essential geometric conditions along with the flare-out condition, thereby supporting asymptotically flat traversable wormholes. To study the underlying matter content responsible for the wormhole structures, we analyze the null energy condition at the wormhole throat and provide graphical representations of various energy conditions, highlighting both the regions where they are satisfied and where they are violated. The stability of the reported wormhole solutions is confirmed through the generalized Tolman–Oppenheimer–Volkoff equation. In addition, we explore several physical features of the wormhole configurations, including the embedding surface, complexity factor, active gravitational mass, and total gravitational energy. Moreover, we investigate the strong deflection of light by these wormholes, finding that the deflection angle approaches zero at large distances, where the wormhole’s gravity is negligible, and diverges near the throat, where the gravitational influence is extremely strong.     
\end{abstract}

\keywords{Dark Matter, Galactic Halo, Kalb-Ramond Gravity, Wormhole, Gravitational Lensing. }

\maketitle

\section{Introduction}
General relativity and various extended theories of gravity allow for the existence of several space-time structures, among which the wormhole is particularly intriguing. This hypothetical tunnel in the fabric of space-time could potentially connect distant regions of the same universe or even different universes. The mysteries of the universe's origin and evolution continue to captivate researchers, and in recent years, traversable wormholes have emerged as a compelling subject in astrophysics, fueled by interest in deep-space travel and the search for extraterrestrial life. The term ’wormhole’ was first introduced by Misner and Wheeler in their seminal research paper entitled ”Classical Physics as Geometry" \cite{cw57}. In fact, the notion of wormholes was originally proposed by Hermann Weyl long ago, in the context of his research aimed at explaining the topological nature of the electromagnetic field \cite{hw21}. To describe a model of a particle in the framework of general relativity without singularities, Einstein and Rosen envisioned space-time as consisting of two identical sheets connected by a region, known as the Einstein-Rosen bridge \cite{Einstein/1935}. Fuller and Wheeler \cite{Fuller/1962} demonstrated that the Einstein-Rosen bridge undergoes instantaneous collapse at the moment of its formation, thereby confirming its non-traversable nature, even for light. The concept of modern traversable Lorentzian wormholes was first introduced by Morris and Thorne in 1988 \cite{ms88}, and later extended by Morris, Thorne, and Yurtsever \cite{ks88}, who demonstrated that sustaining such wormholes within the framework of general relativity requires a violation of the null energy condition (NEC) near the throat of wormhole. Notably, the energy-momentum tensor violating  NEC represents exotic matter. However, their pioneering work presented a static, spherically symmetric metric that connects two asymptotically flat space-times, allowing unrestricted passage of matter and radiation. In this context, Morris and Thorne's traversable wormhole emerges as a well-established solution to Einstein's field equations in general relativity. For a deeper insight into wormhole physics, we refer the reader to the extensive literature on the topic, especially Refs. \cite{mv95, fs74, Ori/1993,Ori/1994,Wang/1995, Galloway/2001,Gao/2000,Graham/2007, Witten/1999,Maldacena/2004,Breitenlohner/1982,Breitenlohner/1982a,Anabalon/2018}.

 The study of wormhole geometries has attracted considerable attention, not only within the framework of Einstein's gravity but also in various modified theories of gravity. Traditionally, general relativity requires exotic matter that violates the standard energy conditions, especially NEC, to support wormhole structures. This has led to significant interest in investigating whether such spacetime configurations can be realised using physically reasonable, non-exotic matter. In the investigation of wormhole geometries, the higher-dimensional gravitational theories \cite{th13, mk13, ka15, rs16, mr19}, including Kaluza-Klein gravity \cite{vd99, jp09, vd14}, higher derivative gravity \cite{hf89,kg92}, nonlocally corrected gravity \cite{ka10}, Dilatonic Einstein-Gauss-Bonnet gravity \cite{pk11},  Randall–Sundrum braneworld gravity \cite{Rs22},  offer significant advantages, most notably, the potential to eliminate the need for exotic matter. This feature has been a major motivation behind the growing interest in modified gravity frameworks. In this broader context, several studies have explored different wormhole solutions in alternative theoretical settings. These include models incorporating scalar fields \cite{cb99}, quantum corrections \cite{sn99}, semiclassical effects \cite{rg07}, and non-singular spacetimes \cite{cb16}. Additionally, wormholes have been studied in brane-world scenarios \cite{la00, ka03} and in the presence of exotic fluids like the Chaplygin gas and its variants \cite{ef12, ef07, ms14, fs06}, Gauss-Bonnet gravity \cite{cg12, sk22}, Rastall gravity \cite{gm22}, Lovelock gravity \cite{mr15, mr21, kc22, kc23}, $f(T)$ gravity \cite{mj13, sr23}, $f(R, T)$ gravity \cite{ab20, zy17}, $f(R)$ gravity \cite{dj23, sc21}, $f(Q)$ gravity \cite{gm21, zh21}, Einstein–Cartan gravity \cite{ns24a}, $\kappa(R,T)$ gravity \cite{ns19a}. Furthermore, Sarif et al. \cite{ms18a} explored the static wormhole solutions in $F (T , T_{G})$ gravity, Sharma et al. \cite{uk22} studied wormhole solutions in the framework of symmetric teleparallel gravity by considering a particular choice of shape and redshift functions.  Fayyaz et al. investigated wormhole structures under the Karmarkar condition \cite{iff20} as well as in logarithmic-corrected $R^2$ gravity \cite{if20}. Shamir et al. \cite{mf21} explored wormhole solutions within the framework of $f(R, G)$ gravity. Gao et al. studied traversable wormholes using a double-trace deformation approach \cite{pg14}, while Visser \cite{vm27} introduced a novel class of traversable wormholes by applying surgical modifications to Schwarzschild spacetimes. Further developments by Maldacena \cite{jm33} explored various theoretical aspects of traversable wormholes.

The Standard Model of Cosmology suggests that approximately 25\% of the universe's total matter content resides in the dark matter (DM) component of the dark sector. Although DM has not yet been directly detected, several theoretical candidates have emerged from particle physics and supersymmetric string theory. Among the most compelling are axions and weakly interacting massive particles (WIMPs), which continue to be actively studied as potential constituents of DM. Astronomer Zwicky was the first to infer the existence of DM in galaxy clusters \cite{zw33, zw37}. Subsequent evidence for DM has emerged from various astrophysical and cosmological observations, including the anomalous behavior of galactic rotation curves \cite{vc80}, galaxy cluster dynamics \cite{fz09}, and the anisotropies in the cosmic microwave background radiation, as measured by the PLANCK satellite \cite{pa16}. Furthermore, several astronomers have provided strong observational evidence for the presence of DM in the Milky Way galaxy \cite{lb04, gb05, pd10}. Notably, Iocco et al. \cite{fi11} offered significant confirmation of DM's existence through detailed analysis. In addition, the studies in Refs. \cite{jd08, vs08} investigated the DM distribution in the galaxy using advanced numerical simulations. Complementing these efforts, Livada et al. \cite{li17} analyzed daily cosmic ray data from the neutron monitor network to deduce variations in cosmic ray density and anisotropy, contributing further insight into galactic DM dynamics. 

The scientific curiosity about wormholes has led to intriguing possibilities regarding their existence in galactic regions, potentially sustained by the presence of DM. Rahaman et al. \cite{frp14} were the first to propose the existence of traversable wormholes within galactic halos, based on the Navarro-Frenk-White (NFW) dark matter density profile. Subsequently, they extended their analysis to suggest the possible presence of wormholes in both the central \cite{fr14} and outer \cite{fr16} regions of galactic halos, employing both the NFW and Universal Rotation Curve (URC) DM density profiles. Jusufi et al. \cite{kj19} investigated the potential formation of wormholes within the galactic halo, attributing their existence to DM in the form of Bose–Einstein condensates. Dark matter-supported wormhole solutions have also been identified in isothermal galactic halos and cosmic voids \cite{ns19}, in the bulge of the Milky Way galaxy \cite{ss20}, as well as within its galactic halo \cite{ss24}.  The modified theories of gravity also contributed to sustaining the wormhole structures in the galactic regions. In this context, wormhole solutions have been proposed within the galactic halo region under the framework of $f(G)$  gravity \cite{ms16}, $f(T)$ gravity \cite{ms14}, $f(Q, T)$ gravity \cite{mt24}, $f(G, T)$ gravity \cite{ms18}, and teleparallel gravity \cite{gm24}, etc.  

In 1936, Einstein \cite{ae36} introduced the concept of gravitational lensing, marking one of the earliest and most profound predictions of general relativity. This phenomenon occurs when light from a distant source is deflected by the gravitational field of a massive object, effectively acting as a lens and providing valuable information about the source. The observational confirmation of light deflection \cite{fw20, as20} significantly amplified interest in gravitational lensing. Since then, its applications have expanded widely, making it an essential tool for exploring exoplanets, DM, and dark energy. A particularly notable regime is strong gravitational lensing, where light can loop multiple times around unstable photon orbits, producing numerous relativistic images \cite{ks00, vb01, vb02}. In recent years, numerous researchers have explored the gravitational lensing effects of various wormhole geometries \cite{pk14, nt16, ns24, wj19, rs17, av18, ns25, kj18, sn20}. Indeed, the study of gravitational lensing in the context of wormholes underscores its growing importance in contemporary wormhole research. 

The preceding discussions inspired us to explore wormhole structures supported by two DM halo models, the King DM density model \cite{ir72} and NFW DM density model \cite{jf96, jf97}, within the framework of Kalb-Ramond gravity. The article is structured as follows: In Sec. \ref{sec2}, we outline the Kalb-Ramond framework for spontaneous Lorentz symmetry breaking. Sec. \ref{sec3} is dedicated to formulating the Einstein field equations in the spacetime of a Morris–Thorne traversable wormhole. Wormhole solutions based on the King dark matter (DM) density model and the Navarro-Frenk-White (NFW) DM density model are presented in Sub-Secs. \ref{sec4a} and \ref{sec4b}, respectively of Sec. \ref{sec4}. In Secs. \ref{sec5} and \ref{sec6}, we examine the various relevant energy conditions and the equilibrium of the proposed models, respectively. Some physical features of the wormhole configurations, including the embedding surface, complexity factor, active gravitational mass, and total gravitational energy, are analyzed in Sec. \ref{sec7}. Sec. \ref{sec9} focuses on the study of light deflection. Finally, the results and concluding remarks are presented in Sec. \ref{sec10}.

 \section{THE KALB-RAMOND FRAMEWORK FOR SPONTANEOUS LORENTZ SYMMETRY BREAKING }\label{sec2}
This section explores the dynamics of the Kalb-Ramond (KR) framework, focusing on the selection and properties of its vacuum expectation value and examining how the resulting framework interacts with gravity. The literature suggests that the KR field can be defined as a rank-2 antisymmetric tensor $B_{\mu\nu}$ in the background of bosonic string theory \cite{mk74}. In differential form, the KR field is described as a 2-form potential $B_2=\frac{1}{2}B_{\mu\nu}dx^\mu\wedge dx^\nu$ with the field strength $H_3=dB_2$, i.e. $H_{\lambda\mu\nu}\equiv\partial_{[\lambda}B_{\mu\nu]}$. Moreover, the KR potential $B_2$ exhibits a gauge symmetry of the form $B_2\rightarrow B_2+d\wedge_1$, where $\wedge_1$ is a 1-form. In some theoretical frameworks, self-interactions of the KR field give rise to configurations that break both the gauge and Lorentz symmetries \cite{va04}. In fact, a self-interacting potential $V = V(B_{\mu\nu}B^{\mu\nu} \pm b_{\mu\nu}b^{\mu\nu})$ with a non-zero vacuum expectation value that meets $\it B_{\mu\nu} \geq b_{\mu\nu}$, establishes a fixed background tensor field in the gravitational sector of the standard model extension, thereby leading to spontaneous breaking of the Lorentz symmetry through Kalb-Ramond self-interaction \cite{va04, ba10}. It should be noted that the potential dependence on $B_{\mu\nu}B^{\mu\nu}$ is preserved invariance under local Lorentz transformations.

 The action for a self-interacting KR field non-minimally coupled to gravity can be expressed as \cite{ba10}.
\begin{equation}
\delta_{KR}^{nonmin}=\int e ~d^4x\left[\frac{R}{2\kappa}-\frac{1}{12}H_{\lambda\mu\nu}H^{\lambda\mu\nu}-V(B_{\mu\nu}B^{\mu\nu}\pm b_{\mu\nu}b^{\mu\nu})+\frac{1}{2\kappa}\left(\xi_2 B^{\lambda\nu}B^\mu_  \nu  R_{\lambda\mu}+\xi_3B^{\mu\nu}B_{\mu\nu}R\right)+\mathcal{L}^M\right],\label{eq2}
\end{equation}

 where $e$ represents the metric determinant, $\kappa=8\pi G$ represents the gravitational coupling constant, and $\xi_3$, $\xi_2$ are the non-minimal coupling constants with dimensions $[\xi]=L^2$. Moreover, the first and second terms correspond to the standard Einstein-Hilbert term and Kalb-Ramond field strength term, respectively, while the third term represents the self-interaction potential responsible for the spontaneous breaking of Lorentz symmetry.
 
\begin{figure}[!htbp]
\begin{center}
\begin{tabular}{rl}
\includegraphics[width=5.9cm]{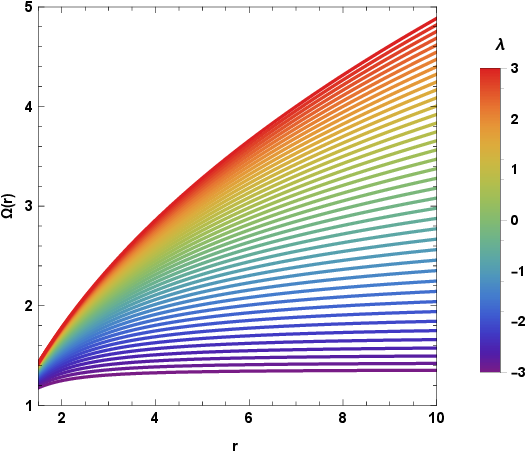}
\includegraphics[width=5.9cm]{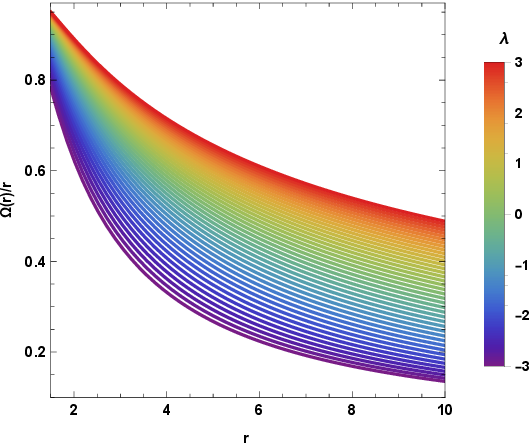}
\includegraphics[width=5.9cm]{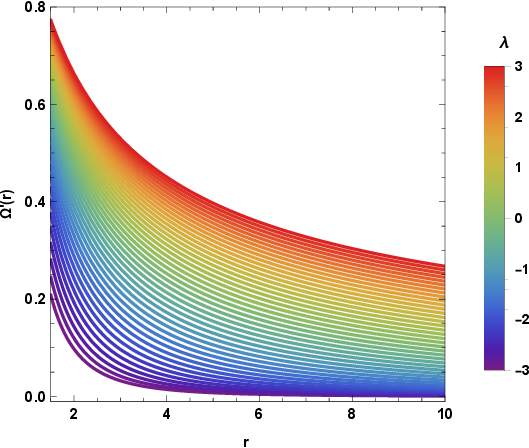}
\\
\end{tabular}
\end{center}
\caption{Shows the radial variations of the shape function $\Omega(r)$ (Left), $\Omega(r)/r$ (Middle), $\Omega'(r)$ (Right) under the King DM model with $\alpha$ = 1.5, $ r_s = 0.3$, $ r_0 = 1$, $\beta$ = 1, and $\eta$ = -3/2.}\label{fig1}
\end{figure}

The present work focuses on the gravitational effects of a wormhole in a phase where the Lorentz symmetry is spontaneously broken. Accordingly, we consider the KR field in its vacuum configuration, characterized by $B_{\mu\nu}B^{\mu\nu} \pm b_{\mu\nu}b^{\mu\nu}$. In fact, the flat spacetime structure permits a constant Lorentz violating vacuum expectation value  $b_{\mu\nu}$, characterized by $\partial_p b_{\mu\nu}=0$ \cite{ba10}. Consequently, the invariant norm $b^2=\eta^{\mu\nu}\eta^{\alpha\beta}b_{\mu\alpha}b_{\nu\beta}$ remains constant, allowing the Lorentz-violating coefficients to be consistently defined across spacetime using $b_{\mu\nu}$ \cite{ba10}. Also, a constant $b_{\mu\nu}$ generates a vanishing KR field strength $h_3=db_2$, therefore, a vanishing vacuum expectation value Hamiltonian \cite{ba10}. A natural extension to curved spacetime is given by $\nabla_\rho b_{\mu\nu}=0$, which ensures that the KR Hamiltonian vanishes \cite{ob05}. Moreover, the KR vacuum expectation value can also be defined by assuming that $b_{\mu\nu}$ has constant norm $b^2=b_{\mu\nu}b^{\mu\nu}$, which is equivalent to $b^{\mu\nu}\nabla_\rho b_{\mu\nu}$ and leads to a vanishing potential \cite{rc18, la20}. Hereafter, we assume a KR vacuum expectation value with a constant norm.
 
We now consider a pseudo-electric configuration in the following form 
\begin{equation}
b_2=-\overset{\sim}{E}(x^1)dx^0\wedge dx^1 .\label{eq3}
\end{equation}
The above Eq. (\ref{eq3}) expresses the vacuum expectation value as $b_2=d \overset{\sim}{A_1}$, where $\overset{\sim}{A_1}=\overset{\sim}{A_0}(x^1)dx^0$ is a pseudo-vector potential and $\overset{\sim}{E}=-\partial_1\overset{\sim}{A_0}$.  Notably, this kind of vacuum expectation value can express a background electric field with a vanishing Hamiltonian. 

 The modified Einstein equations can be obtained by performing a metric variation of Eq. (\ref{eq2}) in the following form \cite{ba10, la20}
\begin{equation}
G_{\mu\nu} = R_{\mu\nu}-\frac{1}{2}R g_{\mu\nu}=\kappa T_{\mu\nu}^{\xi_2}+\kappa T_{\mu\nu}^{M},\label{eq4}
\end{equation}
where $T_{\mu\nu}^{M}$ is the matter field, and
\begin{eqnarray}
T_{\mu\nu}^{\xi_2}&=&\frac{\xi_2}{\kappa}\Big[\frac{1}{2}g_{\mu\nu}B^{\alpha\gamma}B^\beta_ \gamma R_{\alpha\beta}-B^{\alpha}_\mu B^\beta_\nu R_{\alpha\beta}-B^{\alpha\beta}B_{\mu\beta}R_{\nu\alpha}-B^{\alpha\beta}B_{\nu\beta}R_{\mu\alpha}+\frac{1}{2}D_\alpha D_\mu (B_{\nu\beta}B^{\alpha\beta})+\frac{1}{2}D_\alpha D_\nu (B_{\mu\beta}B^{\alpha\beta})\nonumber
\\
&& -\frac{1}{2}D^2 (B^{\alpha}_\mu B_{\alpha\nu})-\frac{1}{2}g_{\mu\nu}D_\alpha D_\beta (B^{\alpha\gamma}B^\beta _\gamma) \Big].\label{eq5}
\end{eqnarray}

We now turn to express the Einstein field equations (\ref{eq4}) for wormhole geometry supported by the stress-energy of Lorentz-violating KR vacuum expectation value (\ref{eq5}).
 
\begin{figure}[!htbp]
\begin{center}
\begin{tabular}{rl}
\includegraphics[width=5.9cm]{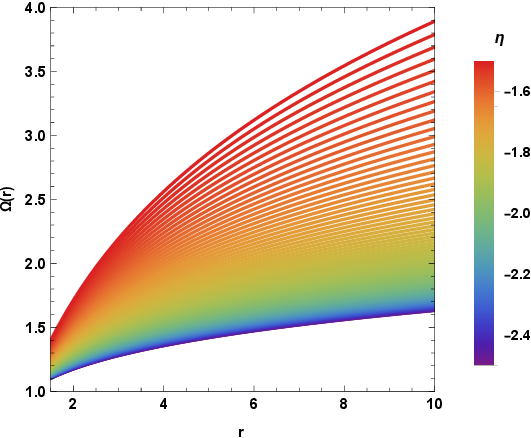}
\includegraphics[width=5.9cm]{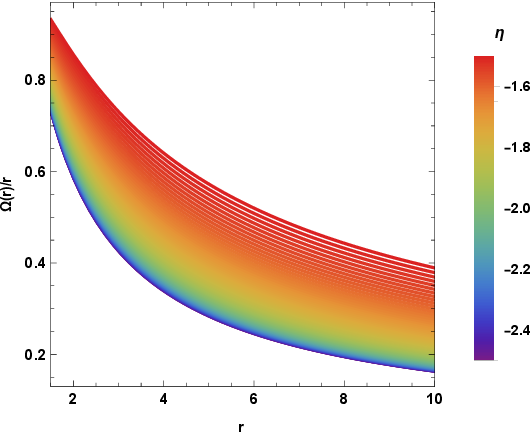}
\includegraphics[width=5.9cm]{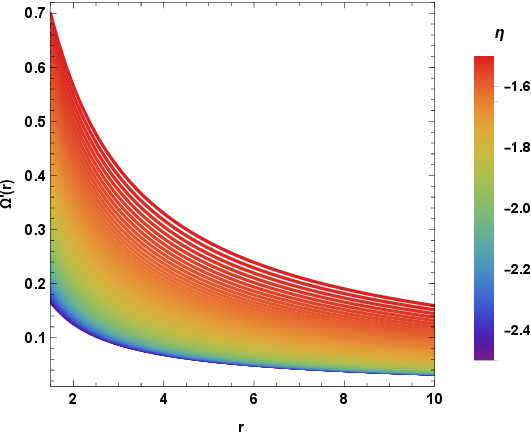}
\\
\end{tabular}
\end{center}
\caption{Shows the radial variations of the shape function $\Omega(r)$ (Left), $\Omega(r)/r$ (Middle), $\Omega'(r)$ (Right) under the King DM model with $\alpha$ = 1.5, $ r_s = 0.3$, $ r_0 = 1$, $\beta$ = 1, and $\lambda$ = 1.}\label{fig2}
\end{figure}

\section{EINSTEIN'S FIELD EQUATIONS}\label{sec3}

To investigate wormhole solutions within the framework of Kalb-Ramond (KR) gravity, we consider the static and spherically symmetric Morris–Thorne traversable wormhole spacetime, defined by \cite{ms88}
\begin{equation}
ds^2=-e^{2\Phi(r)}dt^2+\left(1-\frac{\Omega(r)}{r}\right)^{-1}dr^2+r^2d\theta^2+r^2 sin^2\theta d\phi^2, \label{metric}
\end{equation}
where $\Phi(r)$ and $\Omega(r)$ are termed the redshift function and the shape function of the wormhole, respectively. In particular, the traversable wormhole geometry demands that (i) the redshift function $\Phi(r)$ remains finite throughout the spacetime after the wormhole throat $r = r_0$, where the shape function satisfies $\Omega(r_0) = r_0$, and (ii) the shape function $\Omega(r)$ must satisfy the flare-out condition, $\Omega'(r) < 1$, together with $\Omega(r)/r < 1$ for all $r > r_0$. Here, the prime symbol represents the derivative with respect to the radial coordinate $r$. 

Now, the KR vacuum expectation value ansatz (\ref{eq3}) with the constant norm $b^2=g^{\mu\alpha}g^{\nu\beta}b_{\mu\nu}b_{\alpha\beta}$ yields the following expression for the field component $\overset{\sim}{E}(r)$  in the background of the metric (\ref{metric})
\begin{equation}
\overset{\sim}{E}(r)=\frac{|b|e^{\phi(r)}}{\sqrt{2\left(1-\frac{\Omega(r)}{r}\right)}}, \label{E}
\end{equation}
where $b$ is a constant. It is important to note that the background radial pseudo-electric static configuration $ \overset{\sim}{E^\mu}=(0,\overset{\sim}{E},0,0) $ preserves the spherical symmetry and static nature of the spacetime, as the background vector $\overset{\sim}{E^\mu}$ is orthogonal to both the timelike $t^\mu=(\frac{\partial}{\partial t})^\mu$ and spacelike =$\psi^\mu=(\frac{\partial}{\partial \phi})^\mu$ Killing vectors \cite{ob05}.

For the wormhole metric ansatz (\ref{metric}), the modified Einstein equations (\ref{eq4}) read as follows
\begin{eqnarray}
G_{tt} &=& \frac{\lambda}{4}\left[3R_{tt}-\left(1-\frac{\Omega(r)}{r}\right)R_{rr}\right]+\kappa T_{tt}^M, \label{Gt}\\
G_{rr} &=& \frac{\lambda}{4}\left[3R_{rr}-\left(1-\frac{\Omega(r)}{r}\right)^{-1}R_{tt}\right]+\kappa T_{rr}^M, \label{Gr}\\
G_{\theta\theta} &=& \frac{\lambda r^2}{4}\left[R_{tt}-\left(1-\frac{\Omega(r)}{r}\right)R_{rr}\right]+\kappa T_{\theta\theta}^M, \label{Gth}\\
G_{\phi\phi}  &=& \sin^2\theta G_{\theta\theta},\label{Gph}
\end{eqnarray}
with $\lambda  =|b|^2\xi_2$ as the Kalb-Ramond gravity parameter. In the present study, we now consider the case of zero tidal force, i.e. $\Phi(r) = A = const$.  Consequently, the non-zero components of the Ricci tensor are obtained as
\begin{eqnarray}
R_{rr}=\frac{r\Omega'(r)-\Omega(r)}{r\left(1-\frac{\Omega(r)}{r}\right)},~~~
R_{\theta\theta}=\frac{r\Omega'(r)+\Omega(r)}{2r}, \label{Rth}
\end{eqnarray}

\begin{figure}[!htbp]
\begin{center}
\begin{tabular}{rl}
\includegraphics[width=5.9cm]{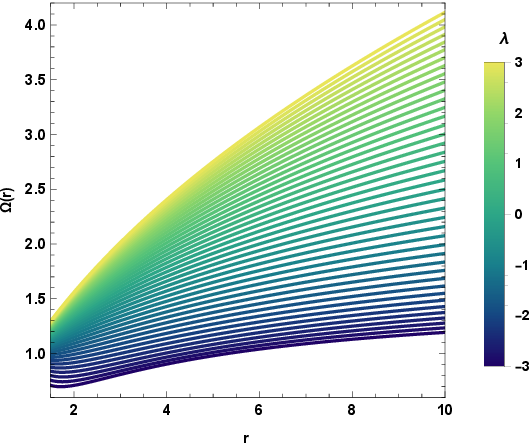}
\includegraphics[width=5.9cm]{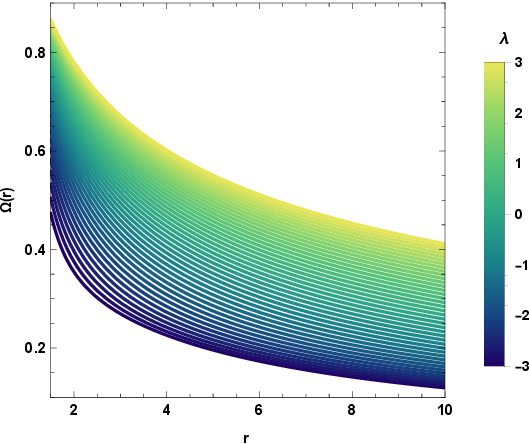}
\includegraphics[width=5.9cm]{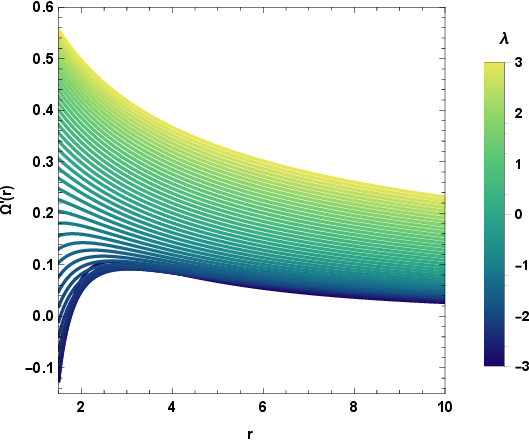}
\\
\end{tabular}
\end{center}
\caption{Shows the radial variations of the shape function $\Omega(r)$ (Left), $\Omega(r)/r$ (Middle), $\Omega'(r)$ (Right) under the NFW DM model with $\rho_s$ = 0.005, $ r_s = 2.1$, $ r_0 = 1$, and $\gamma$ = 2.}\label{fig3}
\end{figure}

In this study, we consider the anisotropic DM matter stress-energy tensor under the background KR field as  
\begin{eqnarray}
    (T^\mu_\nu)^M = \left(\rho(r) +P_r(r)\right)u^\mu u_\nu+P_r(r)g_\mu^\nu + \left(P_t(r)-P_r(r)\right)\chi^\mu \chi_\nu,
\end{eqnarray}
where $u^\mu u_\nu = -\frac{1}{2}\chi^\mu \chi_\nu = -1$. Note that $\rho(r)$, $P_r(r)$, and $P_t(r)$ stand for the energy density, radial pressure, and transverse pressure, respectively. In addition, $u^\mu$ and $\chi^\mu$ are given by $u^\mu = (e^{2\phi(r)},0,0,0)$ and $\chi^\mu = (0,0,1/r,1/\sin\theta)$. Thus, the stress-energy tensors $(T^\mu_\nu)^M$ are obtained as follows
\begin{eqnarray}
    (T^t_t)^M = -\rho(r),~~ (T^r_r)^M = P_r(r),~~ (T^\theta_\theta)^M = (T^\phi_\phi)^M= P_t(r).
\end{eqnarray}
 Therefore, in gravitational units, the modified Einstein field equations (\ref{Gt})-(\ref{Gph}) can be written as
\begin{eqnarray}
\rho(r) &=& \frac{1}{8\pi}\left[\frac{\Omega^\prime(r)}{r^2}+\frac{\lambda}{4r^3}\left(r\Omega^\prime(r)-\Omega(r)\right)\right], \label{den}\\
P_r(r) &=& -\frac{1}{8\pi}\left[\frac{\Omega(r)}{r^3}+\frac{3\lambda}{4r^3}\left(r\Omega^\prime(r)-\Omega(r)\right)\right], \label{pre}\\
P_{t}(r) &=& -\frac{1}{8\pi}\left(1-\frac{\lambda}{2}\right)\frac{r\Omega^\prime(r)-\Omega(r)}{2r^3}.\label{pt}
\end{eqnarray}
In order to construct traversable wormhole solutions one must determine the energy density $\rho(r)$, radial pressure $P_r(r)$, transverse pressure $P_t(r)$, and the shape function $\Omega(r)$ from the above field Eq. (\ref{den})-(\ref{pt}). It is evident that the system (\ref{den})-(\ref{pt}) comprises three equations with four variables, and therefore, we incorporate different DM density profiles to construct the wormhole structures within galactic halos.

\section{WORMHOLE SOLUTIONS INDUCED BY DARK MATTER}\label{sec4}
In this section, we explore the construction of traversable wormhole solutions influenced by DM in the galactic halos under the framework of Kalb-Ramond gravity. In this context, our primary objective is to derive the corresponding shape functions and investigate the essential conditions required for the existence of such wormholes. For this purpose, we consider two well-established DM density models: (i) King’s DM density model and (ii) Navarro-Frenk-White (NFW) DM density model.

\begin{figure}[!htbp]
\begin{center}
\begin{tabular}{rl}
\includegraphics[width=5.9cm]{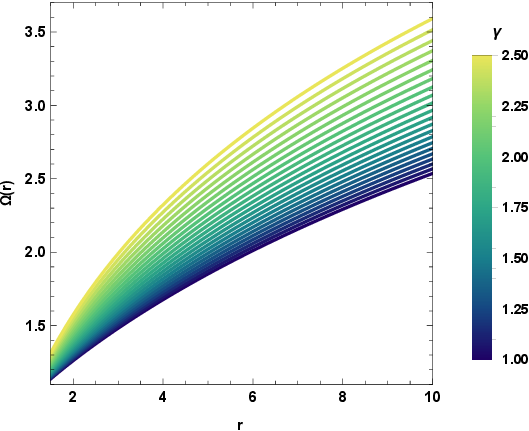}
\includegraphics[width=5.9cm]{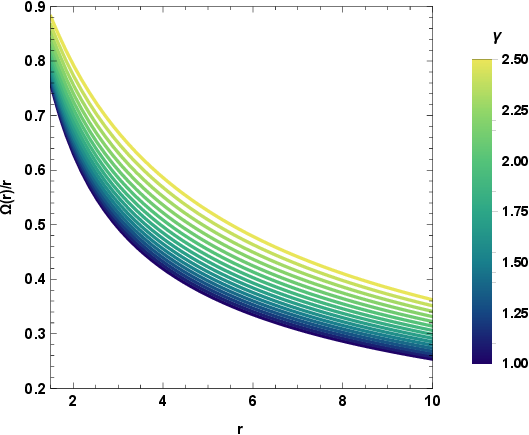}
\includegraphics[width=5.9cm]{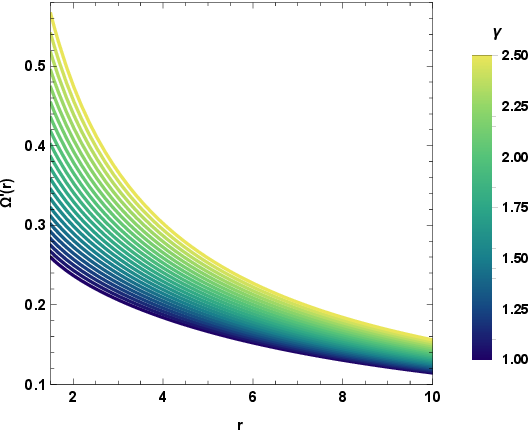}
\\
\end{tabular}
\end{center}
\caption{Shows the radial variations of the shape function $\Omega(r)$ (Left), $\Omega(r)/r$ (Middle), $\Omega'(r)$ (Right) under the NFW DM model with $\rho_s$ = 0.005, $ r_s = 2.1$, $ r_0 = 1$, and $\lambda$ = 1.}\label{fig4}
\end{figure}

\subsection{ King’s Dark Matter Density Model}\label{sec4a}
The King DM density profile is a phenomenological model that is widely used to describe the distribution of DM in galaxies \cite{ir72}.  The literature suggests that the Dragonfly 44 galaxy is identified as an ultra-diffuse galaxy (UDG) with an exceptionally high DM content, comprising over 99\% of its total mass \cite{vd16}. Due to its low surface brightness and flat rotation curves, core-like DM density profiles, such as the King DM model and NFW DM model, etc, are particularly suitable for modelling its DM distribution \cite{an18}. The  King DM density profile is defined by the following expression \cite{ir72, an18}
\begin{equation}
\rho(r)=\alpha\left[\left(\frac{r}{r_s}\right)^2+\beta\right]^\eta,\label{pi}
\end{equation}

where $\alpha$, $\beta$, $\eta$ are parameters, and $r_s$ denotes the
 scale radius. In this study, we assume $\alpha$ = 1.5, $\beta$ = 1 and $\eta \in$ [-5/2,-3/2].

We now derive the corresponding shape function by substituting the density profile (\ref{pi}) into Eq. (\ref{den}), yielding the following expression
\begin{eqnarray}
\Omega(r)&=& \frac{16 \pi  \alpha}{\lambda +6} \beta^\eta H(r) r^3 +C_1 r^{\frac{\lambda }{\lambda +4}}, \label{b}
\end{eqnarray}
where $H(r) = \, _2F_1\left[-\eta ,\frac{\lambda +6}{\lambda +4},\frac{2 (\lambda +5)}{\lambda +4},-\frac{r^2}{r_s^2 \beta }\right]$ is the hypergeometric function and $C_1$ is the integration constant. To determine the integration constant $C_1$, we impose the throat condition $\Omega(r_{0}) = r_{0}$, which gives
\begin{eqnarray}
C_1&=& \frac{r_0^{\frac{4}{\lambda +4}}}{\lambda +6} \left[\lambda +6-16 \pi  \alpha \beta^\eta H(r_0) r_0^2\right].
\end{eqnarray}

Therefore, the shape function (\ref{b}) takes the following form
\begin{eqnarray}
\Omega(r)&=& \frac{1}{\lambda +6}\left[16 \pi  \alpha \beta^\eta H(r) r^3 +r^{\frac{\lambda }{\lambda +4}} r_0^{\frac{4}{\lambda +4}} \left\{\lambda +6-16 \pi  \alpha \beta^\eta H(r_0) r_0^2\right\}\right]. \label{B}
\end{eqnarray}

To ensure an open throat in the wormhole structure, the flaring-out condition must be satisfied. This condition at the throat is expressed by the following relation
\begin{eqnarray}
\Omega'(r_0)&=&  \frac{1}{\lambda +4}\left[\lambda+32 \pi  \alpha  r_0^2 \left(\frac{r_0^2}{r_s^2}+\beta \right)^{\eta }\right]  < 1. \label{b'}
\end{eqnarray}
\begin{figure}[!htbp]
\begin{center}
\begin{tabular}{rl}
\includegraphics[width=5.7cm]{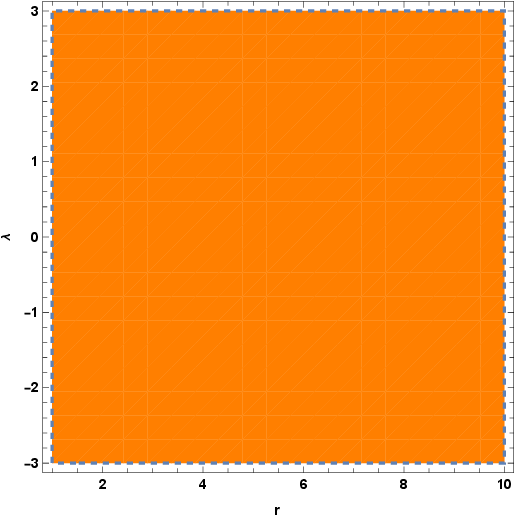}
\includegraphics[width=5.7cm]{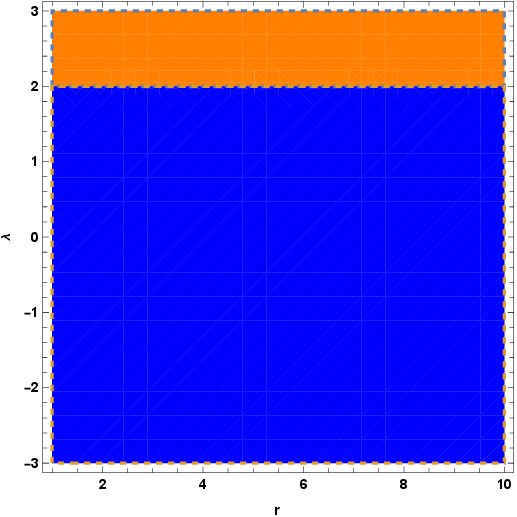}
\includegraphics[width=5.7cm]{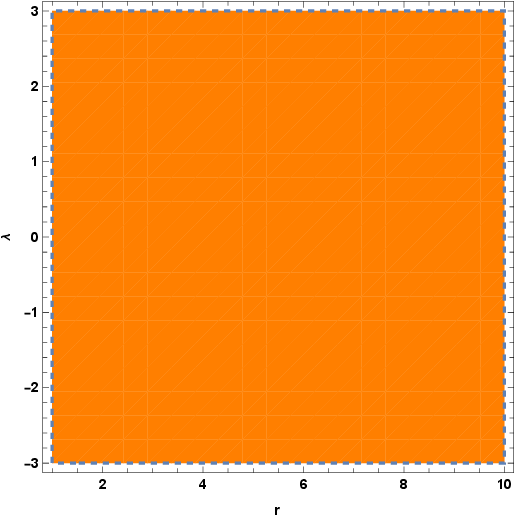}
\\
\end{tabular}
\begin{tabular}{rl}
\includegraphics[width=5.7cm]{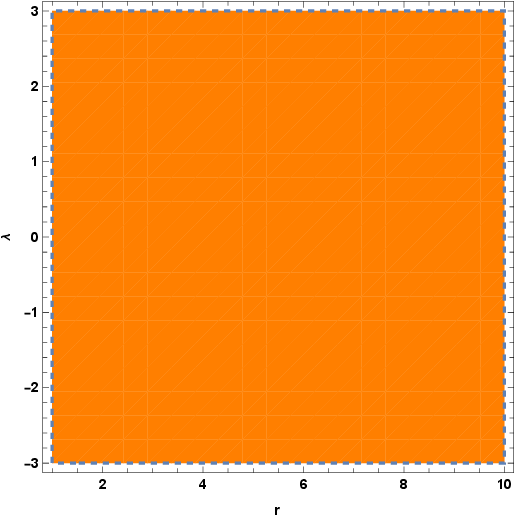}
\includegraphics[width=5.7cm]{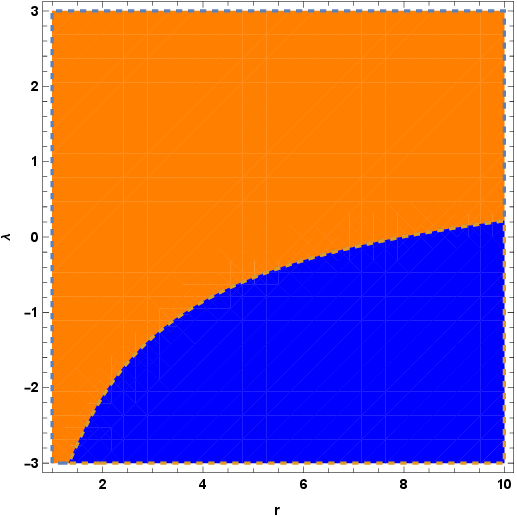}
\includegraphics[width=5.7cm]{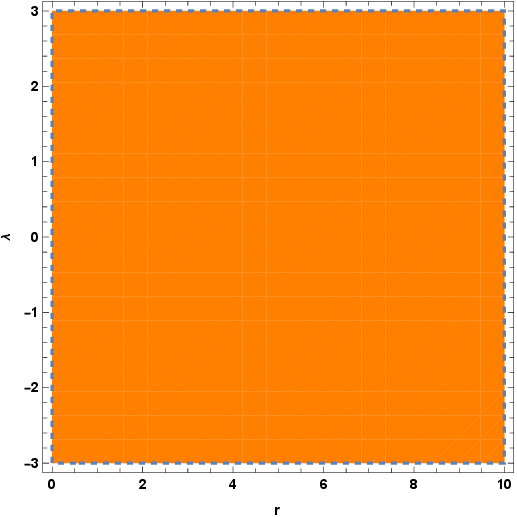}
\\
\end{tabular}
\end{center}
\caption{ Shows the valid region$^{({\color{orange}*})}$ and invalid region$^{({\color{blue}*})}$ for $\rho(r)$ (Left), $\rho(r)+P_r(r)$ (Middle), $\rho(r)+P_t(r)$ (Right) in the above panel, and  $\rho(r)-P_r(r)$ (Left), $\rho(r)-P_t(r)$ (Middle), $\rho(r)+P_r(r)+2P_t(r)$ (Right) in the below panel under the King DM model with $\alpha$ = 1.5, $ r_s = 0.3$, $ r_0 = 1$, $\beta$ = 1, and $\eta$ = -3/2.}\label{fig5}
\end{figure}
 It is worth mentioning that $\lambda\neq$ -4, -6 in this model. To illustrate the behaviours of the obtained shape function (\ref{B}), we present its graphical representation in Fig. \ref{fig1} for the parameter values $\alpha$ = 1.5, $ r_s = 0.3$, $ r_0 = 1$, $\beta$ = 1, $\eta$ = -3/2 and $\lambda \in $[-3, 3] and in Fig. \ref{fig2}  with the parameters $\alpha$ = 1.5, $ r_s = 0.3$, $ r_0 = 1$, $\beta$ = 1, $\lambda$ = 1, and $\eta \in$ [-5/2,-3/2]. Both figures (Fig. \ref{fig1}(Left) and Fig. \ref{fig2}(Left)) indicate that the shape function increases with increasing values of the radial coordinate $r$, the parameter $\lambda$, and the parameter $\eta$. This reveals that the shape function is positively correlated with $r$, $\lambda$, and $\eta$. Moreover, in both cases, the shape function $\Omega(r)$ satisfies $\Omega(r)/r < 1$  for $r > r_0$ (Fig. \ref{fig1}(Middle) and Fig. \ref{fig2}(Middle)) and $\Omega'(r) < 1$  for $r > r_0$ (Fig. \ref{fig1}(Right) and Fig. \ref{fig2}(Right)). Furthermore, Eq. (\ref{b'}) gives $\Omega' (r_0)$ = 0.93968 $<$ 1 for first case with the maximum considered value of $\lambda = 3$, and $\Omega' (r_0)$ = 0.91556 $<$ 1  for second case with the maximum considered value of $\eta = -3/2$. Thus, the obtained shape function satisfies all the essential properties required for constructing the traversable wormhole structures. In addition, one can see from  Figs. \ref{fig1}(Middle) and \ref{fig2}(Middle) that $\Omega(r)/r$ tends to zero as $r$ tends to infinity, therefore, the proposed wormhole solutions are asymptotically flat.
 
\subsection{Navarro-Frenk-White Dark Matter Model}\label{sec4b}

The Cold Dark Matter ($\Lambda$CDM) theory suggests an approximate analytical expression for the Navarro-Frenk-White (NFW) DM density profile, which is further supported by numerical simulations \cite{jf96, jf97}. Here, we adopt the generalized NFW density profile to describe the DM halos, expressed as
\begin{equation}
\rho(r)=\frac{\rho_s}{\left(\frac{r}{r_s}\right)^\gamma\left(1+\frac{r}{r_s}\right)^{3-\gamma}},\label{nfw}
\end{equation}

where the parameter $\rho_s$ corresponds to the dark matter density at the time of halo collapse, $r_s$ denotes the scale radius, and $\gamma$ is the inner slope of the profile. It is important to note that $\gamma = 1$ corresponds to the standard NFW DM density profile.

\begin{figure}[!htbp]
\begin{center}
\begin{tabular}{rl}
\includegraphics[width=5.7cm]{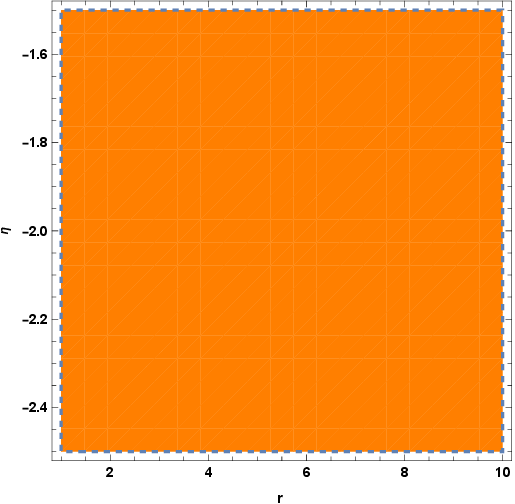}
\includegraphics[width=5.7cm]{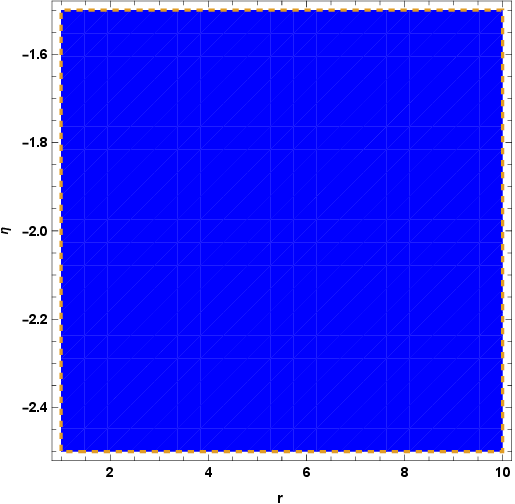}
\includegraphics[width=5.7cm]{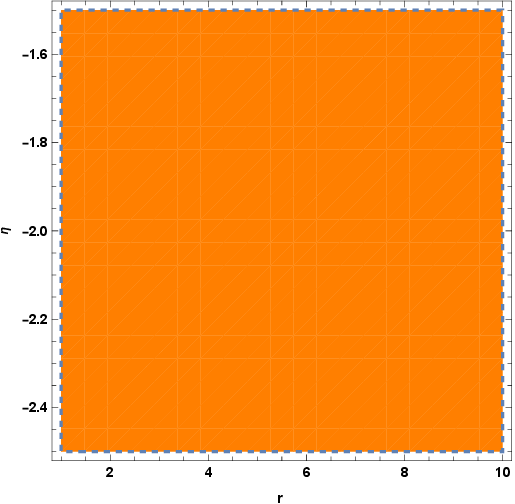}
\\
\end{tabular}
\begin{tabular}{rl}
\includegraphics[width=5.7cm]{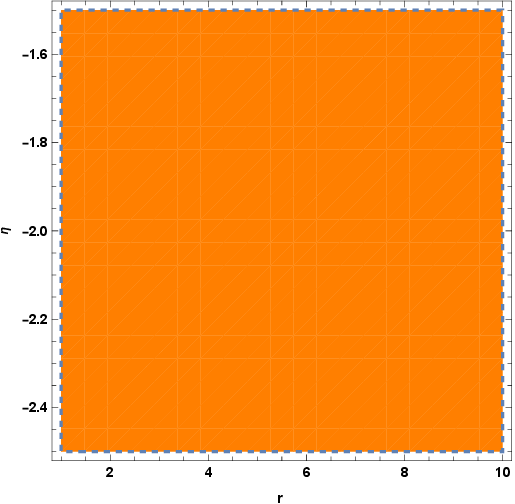}
\includegraphics[width=5.7cm]{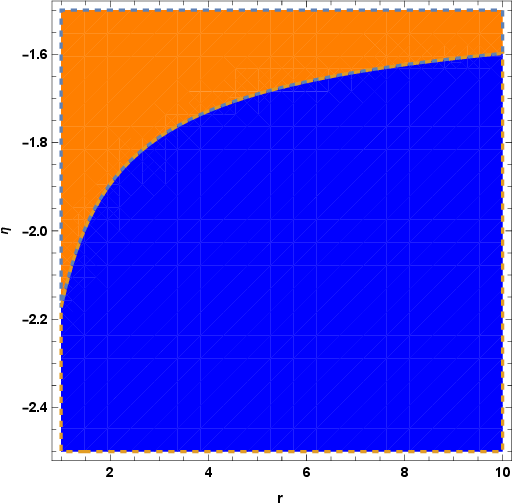}
\includegraphics[width=5.7cm]{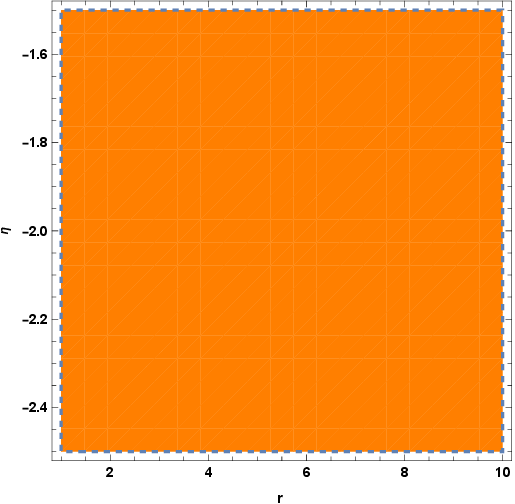}
\\
\end{tabular}
\end{center}
\caption{ Shows the valid region$^{({\color{orange}*})}$ and invalid region$^{({\color{blue}*})}$ for $\rho(r)$ (Left), $\rho(r)+P_r(r)$ (Middle), $\rho(r)+P_t(r)$ (Right) in the above panel, and  $\rho(r)-P_r(r)$ (Left), $\rho(r)-P_t(r)$ (Middle), $\rho(r)+P_r(r)+2P_t(r)$ (Right) in the below panel under the King DM model with $\alpha$ = 1.5, $ r_s = 0.3$, $ r_0 = 1$, $\beta$ = 1, and $\lambda$ = 1.}\label{fig6}
\end{figure}

Substituting the density profile (\ref{nfw}) in  Eq. (\ref{den}), we obtain the following expression for the shape function  
\begin{eqnarray}
\Omega(r)&=& C_2 ((\lambda +4) r)^{\frac{\lambda }{\lambda +4}}-\frac{32 \pi \rho_s G(r) r^3 \left(\frac{r}{r_s}\right)^{-\gamma }}{\gamma  (\lambda +4)-2 (\lambda +6)}, \label{b1}
\end{eqnarray}

where $G(r) = \, _2F_1\left[3-\gamma ,\frac{2 (\lambda +6)}{\lambda +4}-\gamma ,\frac{3 \lambda +16}{\lambda +4}-\gamma ,-\frac{r}{a}\right]$ is a hypergeometric function and $C_2$ is an integration constant determined by applying the throat condition $\Omega(r_{0}) = r_{0}$, taking the following form  
\begin{eqnarray}
C_2&=& \frac{r_0 (r_0 (\lambda +4))^{-\frac{\lambda }{\lambda +4}}}{\gamma  (\lambda +4)-2 (\lambda +6)} \left[\gamma  (\lambda +4)-2 (\lambda +6)+32 \pi  r_0^2\rho_s G(r_0) \left(\frac{r_0}{r_s}\right)^{-\gamma }\right].
\end{eqnarray}

Thus, the shape function (\ref{b1}) becomes in the following form
\begin{eqnarray}
\Omega(r)&=& \frac{r_0 \left( \frac{r}{r_0}\right)^{\frac{\lambda }{\lambda +4}} \left[\gamma  (\lambda +4)-2 (\lambda +6)+32 \pi \rho _s G(r_0) r_0^2 \left(\frac{r_0}{r_s}\right)^{-\gamma }\right]-32 \pi\rho _s  G(r) r^3  \left(\frac{r}{r_s}\right)^{-\gamma }}{\gamma  (\lambda +4)-2 (\lambda +6)}. \label{B1}
\end{eqnarray}

Also, the flaring-out condition at the wormhole throat can be given by the following relation 
\begin{eqnarray}
\Omega'(r_0)&=&  \frac{\lambda  r_0^3+r_s^3 \left[\lambda +32 \pi  r_0^2 \rho _s \left(\frac{r_s+r_0}{r_s}\right)^{\gamma } \left(\frac{r_0}{r_s}\right)^{-\gamma }\right]+3 \lambda  r_0 r_s(r_0+r_s)}{(\lambda +4) \left(r_s+r_0\right){}^3}  < 1. \label{b'1}
\end{eqnarray}

\begin{figure}[!htbp]
\begin{center}
\begin{tabular}{rl}
\includegraphics[width=5.7cm]{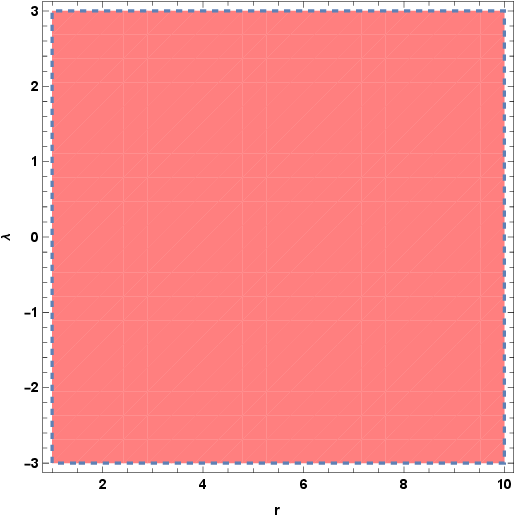}
\includegraphics[width=5.7cm]{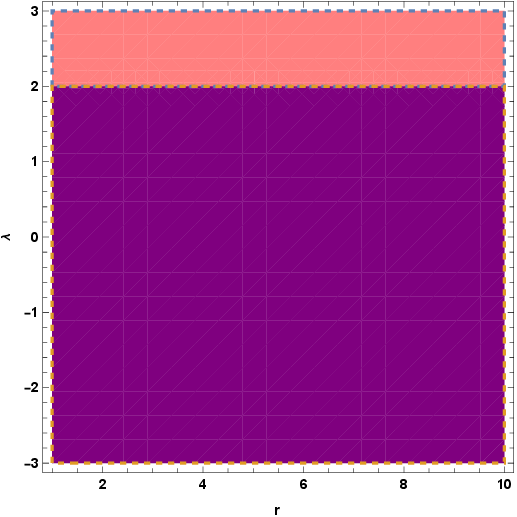}
\includegraphics[width=5.7cm]{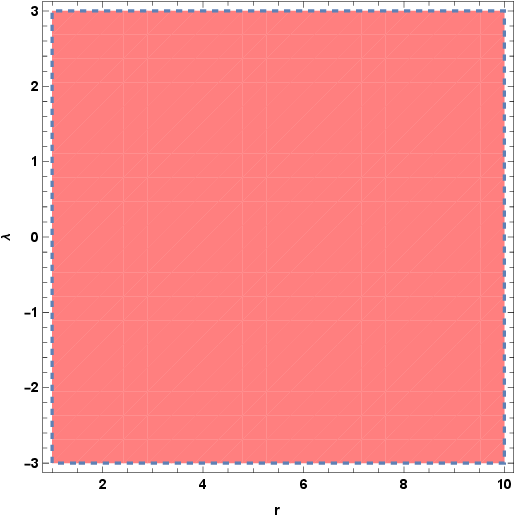}
\\
\end{tabular}
\begin{tabular}{rl}
\includegraphics[width=5.7cm]{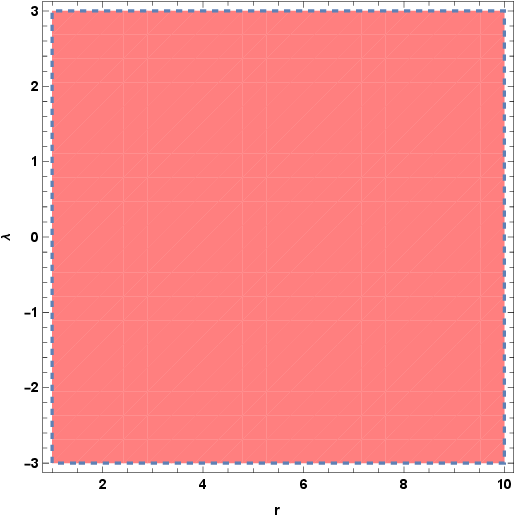}
\includegraphics[width=5.7cm]{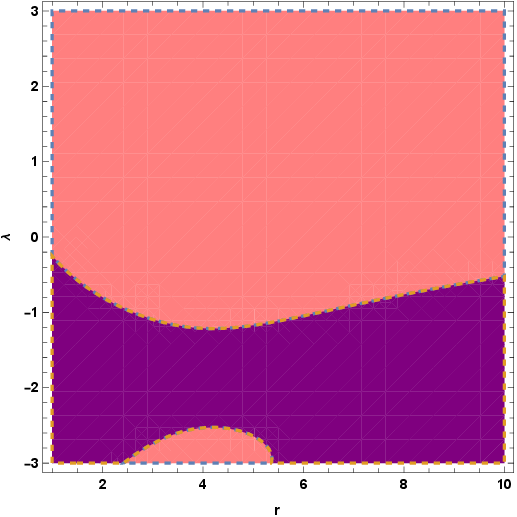}
\includegraphics[width=5.7cm]{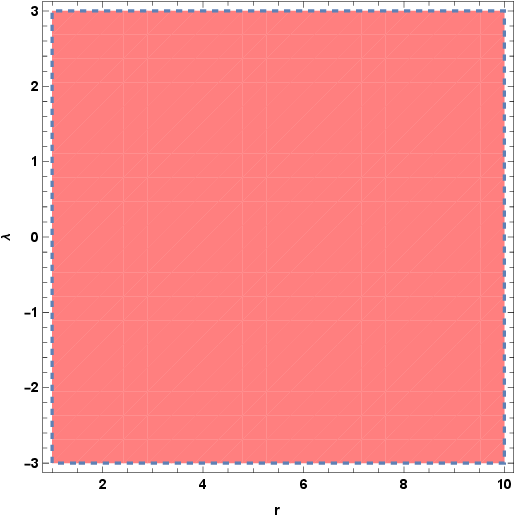}
\\
\end{tabular}
\end{center}
\caption{ Shows the valid region$^{({\color{pink}*})}$ and invalid region$^{({\color{purple}*})}$ for $\rho(r)$ (Left), $\rho(r)+P_r(r)$ (Middle), $\rho(r)+P_t(r)$ (Right) in the above panel, and  $\rho(r)-P_r(r)$ (Left), $\rho(r)-P_t(r)$ (Middle), $\rho(r)+P_r(r)+2P_t(r)$ (Right) in the below panel under the NFW DM model with $\rho_s$ = 0.005, $ r_s = 2.1$, $ r_0 = 1$, and $\gamma$ = 2.}\label{fig7}
\end{figure}

The above result ensures that $\lambda\neq$ - 4 for this model. Now, the obtained shape function (\ref{B1}) is demostrated graphically in Fig. \ref{fig3} corresponding to $\rho_s$ = 0.005, $ r_s = 2.1$, $ r_0 = 1$, $\gamma$ = 2. and $\lambda \in $[-3, 3] and in Fig. \ref{fig4} with $\rho_s$ = 0.005, $ r_s = 2.1$, $ r_0 = 1$, $\lambda$ = 1, and $\gamma \in$ [1, 2.5]. In this model, the shape function exhibits a positive correlation with $\lambda$ and $\gamma$, and similarly with $r$ except in the vicinity of $\lambda = -3$, where the trend reverses near the throat. Also, the shape function $\Omega(r)$ satisfies $\Omega(r)/r < 1$  for $r > r_0$ and $\Omega'(r) < 1$  for $r > r_0$ with $\Omega' (r_0)$ = 0.64309 $<$ 1 for the first case where $\lambda = 3$, and $\Omega' (r_0)$ = 0.72878 $<$ 1  for second case where $\gamma = 2.5$, indicating that the proposed shape function meets of all the necessary conditions to support the wormhole structures. Thus, the shape function derived from the NFW DM density profile meets all criteria for the existence of wormholes in the galactic halos. Moreover, these wormhole solutions are also asymptotically flat in nature.

\section{Energy Condition }\label{sec5}

The literature indicates that ensuring the validity of energy conditions is often a central focus in the study of modified gravitational theories. Indeed, the energy conditions can explain the singularity theorems \cite{sw55} as well as the structure of spacetime. In this context, they play a crucial role in investigating wormhole structures within the framework of modified gravity theories. The primary energy conditions are the null energy condition (NEC), weak energy condition (WEC), dominant energy condition (DEC) and strong energy condition (SEC), all of which are derived from the Raychaudhuri equation \cite{ar55}. The NEC is defined as $T^{\mu\nu} k^\mu k^\nu \geq 0$, .i.e.,
\begin{eqnarray}
    \rho(r) + P_r(r) \geq 0, ~~~~~~~ \rho(r) + P_t(r) \geq 0,
\end{eqnarray}
where $k^\mu$ is the null vector. 

\begin{figure}[!htbp]
\begin{center}
\begin{tabular}{rl}
\includegraphics[width=5.7cm]{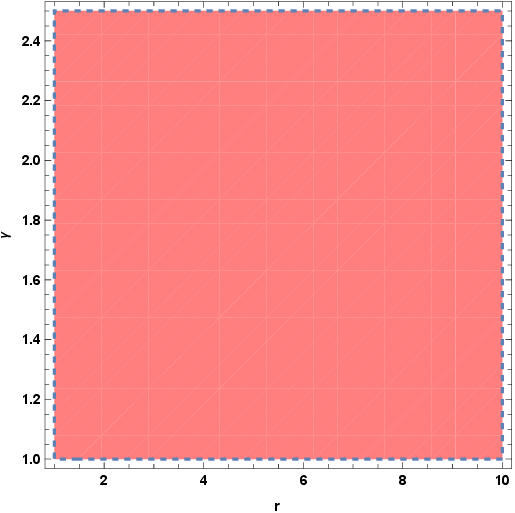}
\includegraphics[width=5.7cm]{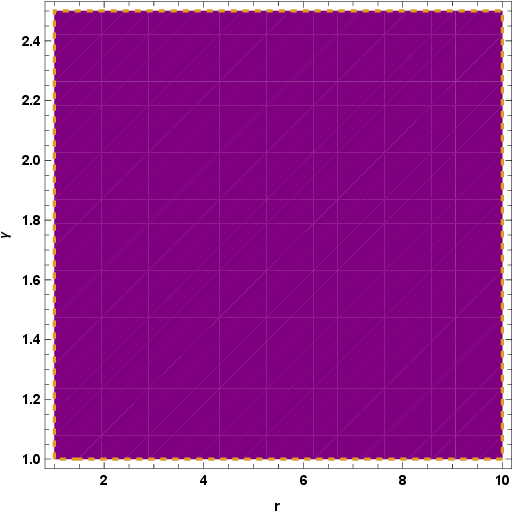}
\includegraphics[width=5.7cm]{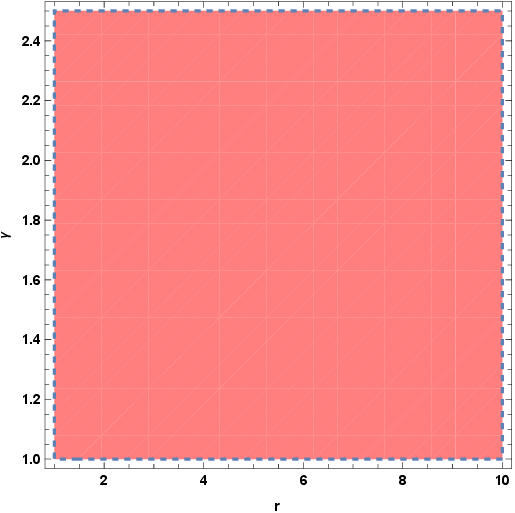}
\\
\end{tabular}
\begin{tabular}{rl}
\includegraphics[width=5.7cm]{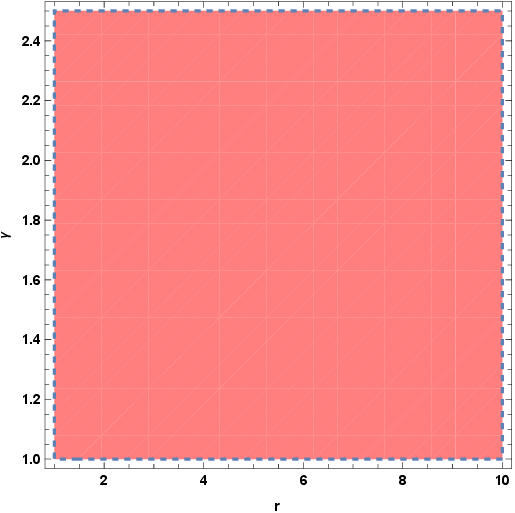}
\includegraphics[width=5.7cm]{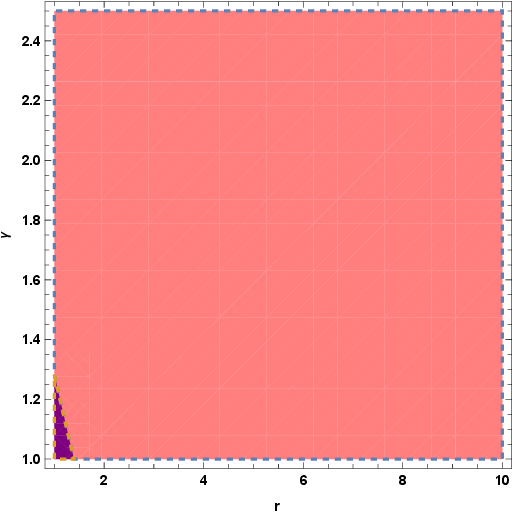}
\includegraphics[width=5.7cm]{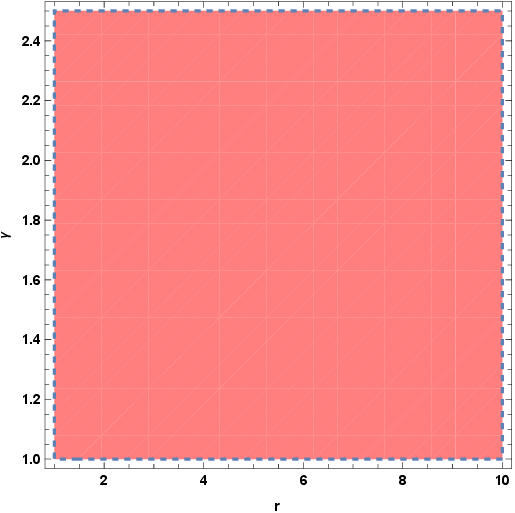}
\\
\end{tabular}
\end{center}
\caption{ Shows the valid region$^{({\color{pink}*})}$ and invalid region$^{( {\color{purple}*})}$ for $\rho(r)$ (Left), $\rho(r)+P_r(r)$ (Middle), $\rho(r)+P_t(r)$ (Right) in the above panel, and  $\rho(r)-P_r(r)$ (Left), $\rho(r)-P_t(r)$ (Middle), $\rho(r)+P_r(r)+2P_t(r)$ (Right) in the below panel under the NFW DM model with $\rho_s$ = 0.005, $ r_s = 2.1$, $ r_0 = 1$, and $\lambda$ = 1.}\label{fig8}
\end{figure}

The WEC is defined as $T^{\mu\nu} U^\mu U^\nu \geq 0$, .i.e.,
\begin{eqnarray}
   \rho(r) \geq 0, ~~~~~~ \rho(r) + P_r(r) \geq 0, ~~~~~~~ \rho(r) + P_t(r) \geq 0,
\end{eqnarray}
where $U^\mu$ is the time-like vector. On the other hand, the DEC requires that
\begin{eqnarray}
    \rho(r) - |P_r(r)| \geq 0, ~~~~~~~ \rho(r) - |P_t(r)| \geq 0,
\end{eqnarray}
and the SEC is defined as
\begin{eqnarray}
    \rho(r) + P_r(r) \geq 0, ~~~~~~~ \rho(r) + P_t(r) \geq 0, ~~~~~~ \rho(r) + P_r(r)+2P_t(r) \geq 0.
\end{eqnarray}

Now, using the above expressions, we examine the behaviour of all energy conditions. In this context, we determine the expression of the radial NEC for the King and NFW models, read as 
\begin{eqnarray}
    \rho(r)+P_r(r) &=&\begin{cases}
\frac{(\lambda -2) \left(\frac{r_0^2}{r_s^2 \beta }+1\right)^{-\eta } \left(\frac{r^2}{r_s^2 \beta }+1\right)^{-\eta }}{4 \pi  (\lambda +4) (\lambda +6) r^3} \bigg[\mathcal{T}_1-16 \pi  \alpha  H(r_0) r_0^{\frac{4}{\lambda +4}+2} r^{\frac{\lambda }{\lambda +4}} \left(\frac{r_0^2}{r_s^2}+\beta \right)^{\eta } \left(\frac{r^2}{r_s^2 \beta }+1\right)^{\eta }\bigg], ~~~\text{(King) }
\\
 \frac{(\lambda -2)\left[32 \pi \rho_s r_0^2 G(r_0)  (r_s+r)^3 (r_0 (\lambda +4))^{\frac{4}{\lambda +4}} ((\lambda +4) r)^{\frac{\lambda }{\lambda +4}} \left(\frac{r_0}{r_s}\right)^{-\gamma }+(\lambda +4) \left\{r_0 \mathcal{T}_2+8 \pi \rho_s  \mathcal{T}_3 r^3 \left(\frac{r}{r_s}\right)^{-\gamma }\right\}\right]}{4 \pi r^3 (\lambda +4)^2 [\gamma  (\lambda +4)-2 (\lambda +6)] (r_s+r)^3}, ~\text{(NFW) }.
\end{cases}
\end{eqnarray}

where
\begin{eqnarray}
        \mathcal{T}_1 &=& \left(\frac{r_0^2}{r_s^2 \beta }+1\right)^{\eta } \Bigg[(\lambda +6) \left(\frac{r^2}{r_s^2 \beta }+1\right)^{\eta } \left\{r_0^{\frac{4}{\lambda +4}} r^{\frac{\lambda }{\lambda +4}}-8 \pi  \alpha  r^3 \left(\frac{r^2}{r_0^2}+\beta \right)^{\eta }\right\}+16 \pi  \alpha  H(r) r^3 \left(\frac{r^2}{r_s^2}+\beta \right)^{\eta }\Bigg],\nonumber
    \\
    \mathcal{T}_2 &=& (r_s+r)^3 [\gamma  (\lambda +4)-2 (\lambda +6)] \left(\frac{r}{r_0}\right)^{\frac{\lambda }{\lambda +4}},\nonumber
    \\
    \mathcal{T}_3&=& r_s^{3-\gamma} [2 (\lambda +6)-\gamma  (\lambda +4)] \left(r_s+r\right)^{\gamma }-4 G(r) (r_s+r)^3.
\end{eqnarray}
\begin{figure}[!htbp]
\begin{center}
\begin{tabular}{rl}
\includegraphics[width=7cm]{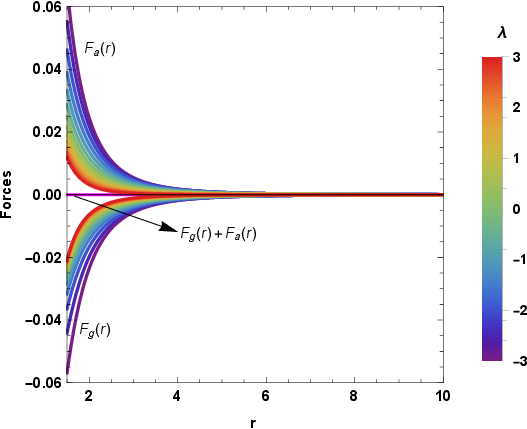}
\includegraphics[width=7.3cm]{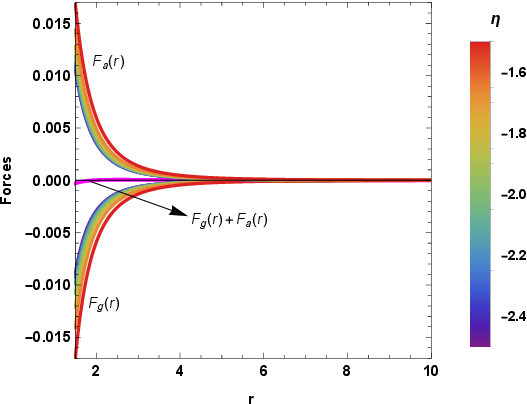}
\\
\end{tabular}
\begin{tabular}{rl}
\includegraphics[width=7cm]{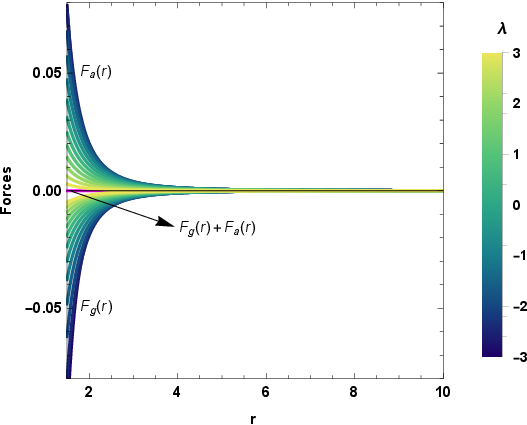}
\includegraphics[width=7.3cm]{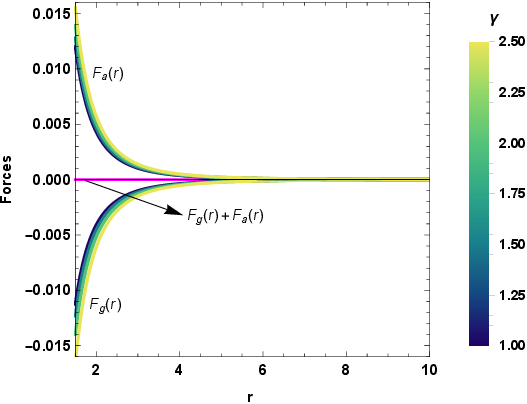}
\\
\end{tabular}
\end{center}
\caption{ Shows the radial variations of the forces under the King DM model with $\alpha$ = 1.5, $ r_s = 0.3$, $ r_0 = 1$, $\beta$ = 1, and $\eta$ = -3/2 (Left), and $\alpha$ = 1.5, $ r_s = 0.3$, $ r_0 = 1$, $\beta$ = 1, and $\lambda$ = 1 (Right) in the above panel; under the NFW DM model with $\rho_s$ = 0.005, $ r_s = 2.1$, $ r_0 = 1$, and $\gamma$ = 2. (Left), and $\rho_s$ = 0.005, $ r_s = 2.1$, $ r_0 = 1$, and $\lambda$ = 1 (Right) in the below panel. }\label{fig9}
\end{figure}

At the throat, the radial NEC for each DM model profile has been derived in the following form
\begin{eqnarray}\label{nec0}
    (\rho(r)+P_r(r))_{r=r_0} &=&\begin{cases}
\frac{(\lambda -2) \left[1-8 \pi  \alpha  r_0^2 \left(\frac{r_0^2}{r_s^2}+\beta \right)^{\eta }\right]}{4 \pi  r_0^2 (\lambda +4)}, ~~~~~~~~~~~~~~~~~~~~~~~~~~~~~\text{(King) }
\\
\frac{(\lambda -2) \left[3 r_s^2 r_0+3 r_s r_0^2+r_0^3+r_s^3 \left\{1-8 \pi \rho_s r_0^2  \left(\frac{r_0}{r_s+r_0}\right)^{-\gamma }\right\}\right]}{4 \pi  r_0^2 (\lambda +4) (r_s+r_0)^3}, ~\text{(NFW) }.
\end{cases}
\end{eqnarray}

\begin{figure}[!htbp]
\begin{center}
\begin{tabular}{rl}
\includegraphics[width=5.5cm]{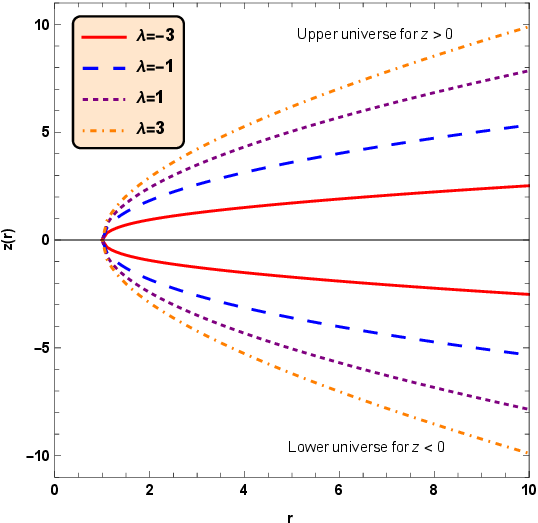}
\includegraphics[width=5.5cm]{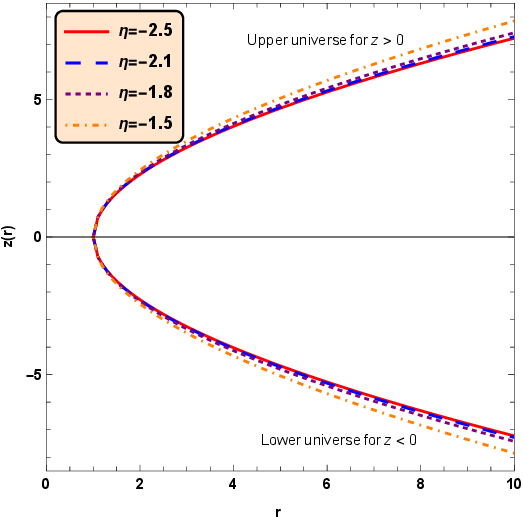}
\includegraphics[width=7.8cm]{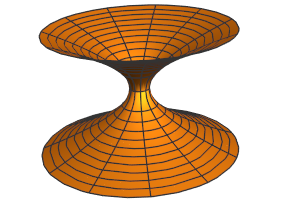}
\\
\end{tabular}
\end{center}
\caption{ Shows the radial variations of the embedding surface under the King DM model with $\alpha$ = 1.5, $ r_s = 0.3$, $ r_0 = 1$, $\beta$ = 1, and $\eta$ = -3/2 (Left), and $\alpha$ = 1.5, $ r_s = 0.3$, $ r_0 = 1$, $\beta$ = 1, and $\lambda$ = 1 (Middle), the full visualization diagrams of wormhole with $ \alpha = 1.5$, $ \beta = 1$, $r_s = 0.3$, $r_0 = 1$, $\lambda = 1$, and $\eta = -3/2$ (Right).}\label{fig10}
\end{figure}

\subsection{Analysis of Energy Conditions}

Here, we aim to analyze all the energy conditions for both proposed wormhole models. This comprehensive investigation will help to assess the nature of the matter threading the wormhole solutions and determine the conditions under which exotic or nonexotic matter is present in the context of Kalb-Ramond gravity. The result (\ref{nec0}) indicates that $ (\rho(r)+P_r(r))_{r=r_0} = 0$ for both King's and NFW models when the parameter $\lambda = 2$, thereby ensuring that the radial NEC is marginally satisfied at the throat for $\lambda = 2$. However, to fully characterize the matter content, it is essential to examine all the relevant energy conditions as they collectively confirm the nature of matter content at the throat of the wormholes. More broadly, research into wormhole structures and their corresponding energy conditions is vital for uncovering the theoretical foundations and physical implications of these intriguing spacetime entities. In this context, the extensive graphical analyses that highlight the regional behaviours of the energy conditions in the framework of the considered DM models are illustrated in Figs. \ref{fig5}-\ref{fig8}. In fact, these analyses reveal the valid regions where the energy conditions are satisfied and the invalid regions where the energy conditions are violated, depending on the range of model parameters $\lambda$, $\eta$ and $\gamma$,  highlighting the critical role of these parameters. The energy density $\rho(r)$ for the King model exhibits positive behaviour throughout the entire spacetime in both the cases $-3 \leq \lambda \leq 3$ (See Fig. \ref{fig5}) and $-5/2 \leq \eta \leq -3/2$ (See Fig. \ref{fig6}). The energy density $\rho(r)$ also shows the positive behaviour for the NFW model with $-3 \leq \lambda \leq 3$ (See Fig. \ref{fig7}) and $1 \leq \gamma \leq 2.5$ (See Fig. \ref{fig8}). For the King model,  the radial NEC $\rho(r)+P_r(r)$ is violated for $-3 \leq \lambda < 2$, and satisfied for $2 \leq \lambda \leq 3$ in the first case, whereas it is completely violated in the second case where  $-5/2 \leq \eta \leq -3/2$. The invalid region of $\rho(r)+P_r(r)$ for the NFW model is $-3 \leq \lambda < 2$, and the valid region is $2 \leq \lambda \leq 3$ in the first case, whereas $1 \leq \gamma \leq 2.5$ is the invalid region in the second case throughout the entire spacetime. For both the models, $\rho(r)+P_t(r)$, $\rho(r)-P_r(r)$, and $\rho(r)+P_r(r)+2P_t(r)$ are positive within the specified rage of parameters. However, in the first scenario of the king model, the quantity $\rho(r)-P_t(r)$ remains valid within the range $-3 \leq \lambda \leq 3$ for $1\leq r \leq 1.4$. Beyond this, for $1.4 < r \leq 10$, the valid region gradually decreases while the invalid region expands, reaching up to $-3 \leq \lambda \leq 0.19$. In the second scenario of the king model, $\rho(r)-P_t(r)$ has the maximum invalid region up to $-5/2 \leq \lambda \leq -1.6$. For the NFW model, $\rho(r)-P_t(r)$ shows the maximum valid regions in both cases. It is noteworthy that all the model parameters $\lambda$, $\eta$, and $\gamma$ play a crucial role in influencing all the energy conditions in the present study. All the energy conditions except NEC are predominantly satisfied within the considered parameter range. The scenario of the radial NEC for both models reveals that the Kalb-Ramond gravity parameter $\lambda$ significantly influences the nature of the matter content, indicating the existence of exotic matter for $-3 \leq \lambda < 2$ and nonexotic matter for $2 \leq \lambda \leq 3$. The exotic matter is believed to contribute to the stability and traversability of wormholes by counterbalancing the gravitational collapse typically caused by ordinary matter. Consequently, the Kalb-Ramond gravity framework supports the wormhole solutions with the presence of exotic as well as nonexotic matter, depending on the values of the Kalb-Ramond gravity parameter $\lambda$. This dual possibility highlights the flexibility of the Kalb-Ramond gravity theory in accommodating a broader range of matter content while admitting traversable wormhole solutions. A summary of all the aforementioned results concerning the energy conditions for each DM model is provided in Tables-\ref{tab1} and \ref{tab2}.

\section{Equilibrium Analysis}\label{sec6}
In this section, we employ the generalized Tolman–Oppenheimer–Volkoff (TOV) equation to assess the stability of the wormhole solutions obtained. The generalized TOV equation is given by the following expression \cite{jr39}
\begin{eqnarray}
	&&\frac{\omega'}{2}[\rho(r)+P_r(r)]+\frac{dP_r(r)}{dr}+\frac{2}{r}[P_r(r)-P_t(r)]=0.\label{t}
\end{eqnarray}
In the present study, $\omega = 2\Phi = 2A$, constant. Due to the anisotropic distribution of matter, we define the gravitational force, hydrostatic force, and anisotropic force as follows
\begin{eqnarray}
    F_g(r) = - \frac{\omega'}{2}[\rho(r)+P_r(r)],~~ F_h(r) = -dP_r(r)/dr,~~ F_a(r) = \frac{2}{r}[P_t(r)-P_r(r)].
\end{eqnarray}

Thus, the equilibrium of the wormhole solutions is governed by the following key condition 
\begin{eqnarray}
F_g(r)+F_h(r)+F_a(r)=0. 
\end{eqnarray}

In the present study,  $ F_g(r) = 0$ due to the choice of the constant redshift function. The hydrostatic force $F_h(r)$, and anisotropic force $F_a(r)$ are graphically represented in the above panel of Fig. \ref{fig9} for the King model and in the below panel of Fig. \ref{fig9} for the NEW model. These figures confirm that the wormhole configurations are in equilibrium under the combined influence of hydrostatic and anisotropic forces.

\begin{figure}[!htbp]
\begin{center}
\begin{tabular}{rl}
\includegraphics[width=5.5cm]{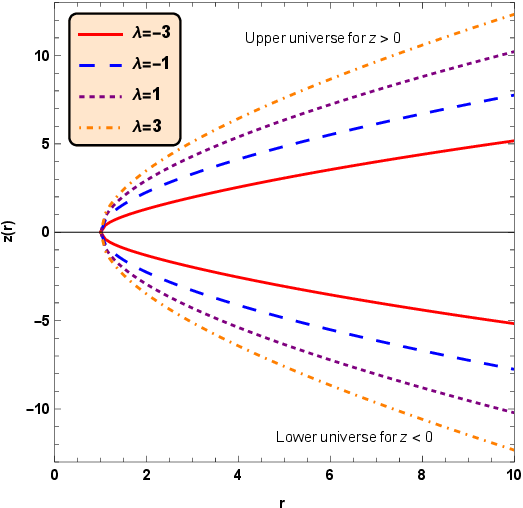}
\includegraphics[width=5.5cm]{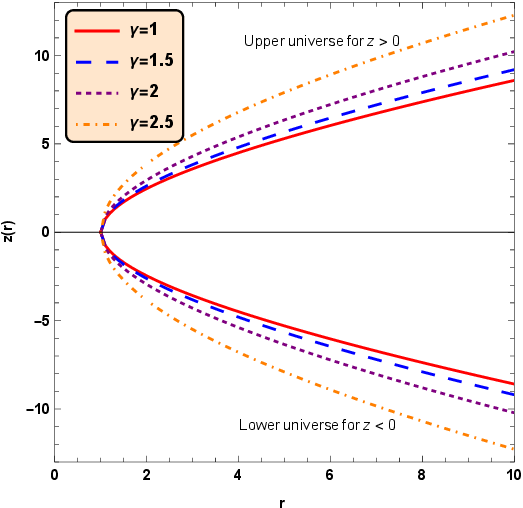}
\includegraphics[width=7cm]{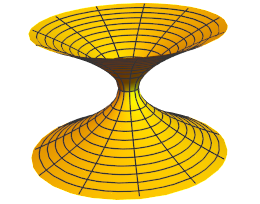}
\\
\end{tabular}
\end{center}
\caption{ Shows the radial variations of the embedding surface under the NFW DM model with $\rho_s$ = 0.005, $ r_s = 2.1$, $ r_0 = 1$, and $\gamma$ = 2. (Left), and $\rho_s$ = 0.005, $ r_s = 2.1$, $ r_0 = 1$, and $\lambda$ = 1 (Middle), the full visualization diagrams of wormhole with $ \rho_s = 0.005$,  $r_s = 2.1$, $r_0 = 1$, $\lambda = 1$, and $\gamma = 2$ (Right).}\label{fig11}
\end{figure}

\section{Some Physical Features of Wormholes}\label{sec7}

\subsection{Embedding Surface}\label{sec7a}

This part of the study focuses on employing embedding diagrams to understand the geometric structure of wormhole spacetime better.  In this context, our main focus is on the geometry of wormholes, which requires specific constraints on the choice of coordinates. Thus,  we consider an equatorial slice $\theta = \pi/2$ with a fixed time $t = constant$, which yields the following result for the wormhole metric (\ref{metric}) 
\begin{equation}
ds^2 = -\left(1-\frac{\Omega(r)}{r}\right)^{-1}dr^2-r^2d\phi^2,\label{es}
\end{equation}
Now, the embedded surface $z(r)$ of the axially symmetric wormhole in three-dimensional space is expressed as \cite{ms88}
\begin{equation}
ds^2 = -\left[1+\left(\frac{d z(r)}{dr}\right)^2\right]dr^2-r^2 d\phi^2.\label{es1}
\end{equation}
Thus, one can obtain the following differential equation of the embedding surface $z(r)$ from Eqs. (\ref{es}) and (\ref{es1})  
\begin{equation}
\frac{dz(r)}{dr} = \pm\left(\frac{r}{\Omega(r)}-1\right)^{-\frac{1}{2}}.\label{dz1}
\end{equation}
The above Eq. (\ref{dz1}) shows that $dz(r)/dr$ diverges at the wormhole throat, and therefore, $z(r)$  becomes vertical there. Furthermore, as $r \to \infty$, $dz/dr$ approaches zero, indicating that the spacetime flattens far from the throat. Now, Eq. (\ref{dz1}) gives the integral form of the embedding surface $z(r)$ as
\begin{equation}
z(r) = \pm \int_{r_0^+}^\infty \left(\frac{r}{\Omega(r)}-1\right)^{-\frac{1}{2}}.\label{z1}
\end{equation}
Here, the $\pm$ sign corresponds to the upper and lower universes of the wormhole geometry, respectively. By substituting the shape function (\ref{B}) for the King model, and the shape function (\ref{B1}) for the NFW model into Eq. (\ref{z1}), the corresponding embedding surface and the complete wormhole visualisation diagram are shown in Figs. \ref{fig10} and \ref{fig11}, respectively. In these diagrams, positive curvature ($z > 0$) corresponds to the upper universe, while negative curvature ($z < 0$) represents the lower universe.

\begin{figure}[!htbp]
\begin{center}
\begin{tabular}{rl}
\includegraphics[width=7cm]{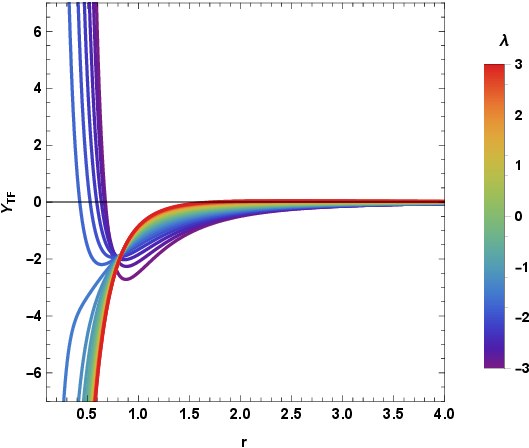}
\includegraphics[width=7.3cm]{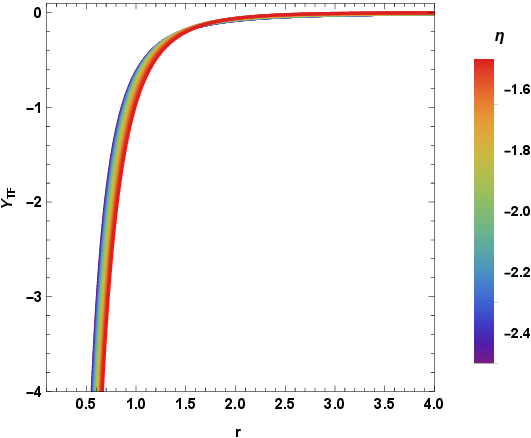}
\\
\end{tabular}
\begin{tabular}{rl}
\includegraphics[width=7cm]{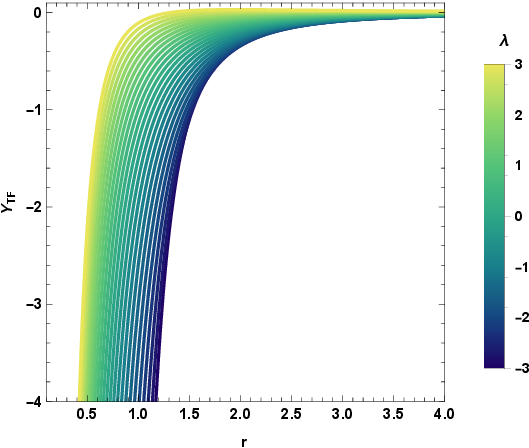}
\includegraphics[width=7.3cm]{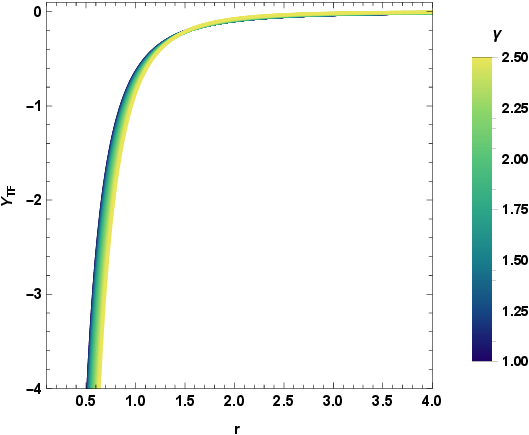}
\\
\end{tabular}
\end{center}
\caption{ Shows the radial variations of the complexity factor under the King DM model with $\alpha$ = 1.5, $ r_s = 0.3$, $ r_0 = 1$, $\beta$ = 1, and $\eta$ = -3/2 (Left), and $\alpha$ = 1.5, $ r_s = 0.3$, $ r_0 = 1$, $\beta$ = 1, and $\lambda$ = 1 (Right) in the above panel; under the NFW DM model with $\rho_s$ = 0.005, $ r_s = 2.1$, $ r_0 = 1$, and $\gamma$ = 2 (Left), and $\rho_s$ = 0.005, $ r_s = 2.1$, $ r_0 = 1$, and $\lambda$ = 1 (Right) in the below panel.}\label{fig12}
\end{figure}

\subsection{ Complexity Factor}\label{sec7b}
Herrera \cite{lh18} proposed a suitable definition of complexity for self-gravitating matter configurations, stating that the least complexity of a gravitational system can be characterized by homogeneous energy density and isotropic pressure. Furthermore, the literature suggests that the associated scalar in static spherically symmetric spacetimes can be derived through the orthogonal splitting of the Riemann tensor \cite{lh18, ag08, lh09}. In fact, this scalar encapsulates the essential features of complexity, it is denoted by $Y_{TF}$ and is expressed as follows
\begin{eqnarray}
    Y_{TF} = 8\pi \Delta(r)-\frac{4\pi}{r^3}\int_0^r \bar{r}^3\rho'(\bar{r})d\bar{r},\label{ytf}
\end{eqnarray}
where $\Delta(r) = P_r(r)-P_t(r)$. From the above result (\ref{ytf}), one can express the Tolman mass $m_T$  in the following form 
\begin{eqnarray}
    m_T = (m_T)_R\left(\frac{r}{R}\right)^3 +r^3 \int_0^{R}\frac{e^{(\nu+\xi)/2}}{\bar{r}}Y_{TF}d\bar{r}.
\end{eqnarray}
It is important to note that the above result fully reflects all the influences arising from both energy density inhomogeneity and pressure anisotropy on the active gravitational mass, thus providing a solid basis for defining the complexity factor in terms of the scalar (\ref{ytf}). The vanishing complexity condition $Y_{TF} = 0$ yeilds
\begin{eqnarray}
  \Delta(r)=\frac{1}{2r^3}\int_0^r \bar{r}^3\rho'(\bar{r})d\bar{r}.\label{ytf1}
\end{eqnarray}
The above condition can be satisfied not only in the simplest case of isotropic and homogeneous systems but also in a wide range of other scenarios. In this context, satisfying the vanishing complexity condition leads to a non-local equation of state, which serves as an additional relation to close the system of Einstein’s field equations. This approach has been explored in several recent studies \cite{rc09, ec19, ca22}. Here, we are willing to study the complexity factor for the proposed wormhole solutions in galactic regions. For the wormhole geometry with a finite size of wormhole throat $r_0$, the complexity factor (\ref{ytf}) can be expressed as
\begin{eqnarray}
    Y_{TF} = 8\pi \Delta(r)-\frac{4\pi}{r^3}\int_{r_0}^r \bar{r}^3\rho'(\bar{r})d\bar{r}.\label{ytf1}
\end{eqnarray}

\begin{figure}[!htbp]
\begin{center}
\begin{tabular}{rl}
\includegraphics[width=7.3cm]{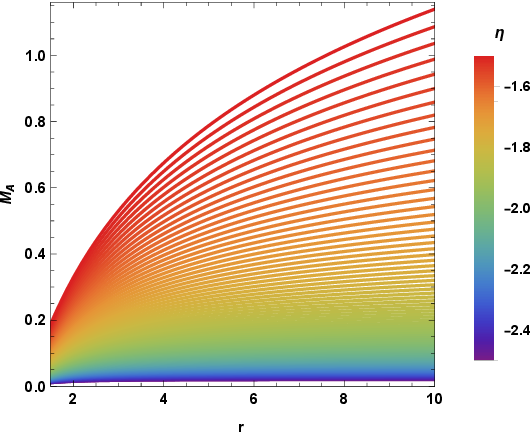}
\includegraphics[width=7.3cm]{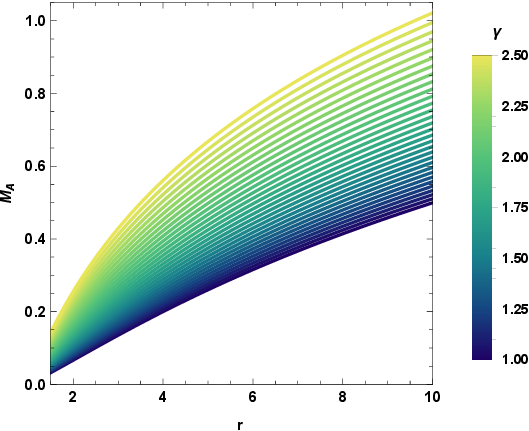}
\\
\end{tabular}
\end{center}
\caption{ Shows the radial variations of the active gravitational mass under the King DM model with $\alpha$ = 1.5, $ r_s = 0.3$, and $\beta$ = 1 (Left); under the NFW DM model with $\rho_s$ = 0.005, and $ r_s = 2.1$ (Right).}\label{fig13}
\end{figure}
Now, based on Eq. (\ref{ytf1}), we have graphically analyzed the behaviour of the complexity factor for the present wormhole models, as shown in Fig. \ref{fig12}. It is found that the complexity factor for the present wormhole models tends to zero as $r \rightarrow \infty$. Consequently, for the present galactic halo wormhole models in Kalb-Ramond gravity, the complexity factor tends to zero as the radial distance moves far away from the throat of the wormhole. Notably, a minimal complexity factor is associated with a configuration exhibiting homogeneous energy density and isotropic pressure \cite{rc09}. Subsequently, a vanishing complexity factor occurs in the configurations with inhomogeneous energy density and anisotropic pressure, as long as these two effects exactly offset each other. In this context, anisotropic pressure plays a more dominant role than energy density homogeneity in the complexity factor \cite{ms18, zy20, mz23}.

\begin{table}[thth]
\caption{ The behaviour of the energy conditions for the King DM model.}\label{tab1}
\centering
\begin{tabular}{|c|c|c|c|c|c|c|c|cc}
\hline
\multicolumn{2}{|c|}{For $\alpha$ = 1.5, $ r_s = 0.3$, $ r_0 = 1$, $\beta$ = 1, and $\eta$ = -3/2}\\
\hline
Expressions  & Results \\
\hline
 $\rho(r)$ (Fig. \ref{fig5} left above panel (AP)) & $\rho(r) > 0$ for $-3 \leq \lambda \leq 3$ within $r\geq r_0$ \\
\hline
 $\rho(r) + P_r(r)$ (Fig. \ref{fig5} middle AP) &  $\rho(r) + P_r(r) < 0$ for $-3 \leq \lambda<2$; $\rho(r) + P_r(r) \geq 0$ for $2 \leq \lambda \leq 3$  within $r\geq r_0$  \\
\hline
 $\rho(r) + P_t(r)$ (Fig. \ref{fig5} right AP) &  $\rho(r) + P_t(r) > 0$ for $-3 \leq \lambda \leq 3$ within $r\geq r_0$ \\
\hline
 $\rho(r) - P_r(r)$  (Fig. \ref{fig5} left below panel (BP)) &  $\rho(r) - P_r(r) > 0$ for $-3 \leq \lambda \leq 3$  within $r\geq r_0$   \\
\hline
 $\rho(r) - P_t(r)$ (Fig. \ref{fig5} middle BP) &  $\rho(r) - P_t(r) > 0$ for $-3 \leq \lambda \leq 3$  within $r_0\leq r < 1.4$; then (-)ve region expands   \\
\hline
 $\rho(r) + P_r(r) + 2P_t(r)$ (Fig. \ref{fig5} right BP) &   $\rho(r) + P_r(r) + 2P_t(r) > 0$ for $-3 \leq \lambda \leq 3$  within $r\geq r_0$    \\
\hline
\hline
\multicolumn{2}{|c|}{For $\alpha$ = 1.5, $ r_s = 0.3$, $ r_0 = 1$, $\beta$ = 1, and $\lambda$ = 1}\\
\hline
Expressions  & Results \\
\hline
 $\rho(r)$ (Fig. \ref{fig6} left above panel (AP)) & $\rho(r) > 0$ for $-5/2 \leq \eta  \leq -3/2$ within $r\geq r_0$ \\
\hline
 $\rho(r) + P_r(r)$ (Fig. \ref{fig6} middle AP) &  $\rho(r) + P_r(r) < 0$ for $-5/2 \leq \eta  \leq -3/2$ within $r\geq r_0$  \\
\hline
 $\rho(r) + P_t(r)$ (Fig. \ref{fig6} right AP) &  $\rho(r) + P_t(r) > 0$ for $-5/2 \leq \eta  \leq -3/2$ within $r\geq r_0$ \\
\hline
 $\rho(r) - P_r(r)$  (Fig. \ref{fig6} left below panel (BP)) &  $\rho(r) - P_r(r) > 0$ for $-5/2 \leq \eta  \leq -3/2$ within $r\geq r_0$   \\
\hline
 $\rho(r) - P_t(r)$ (Fig. \ref{fig6} middle BP) &  $\rho(r) - P_t(r)$ has maximum (-)ve region up to $-5/2 \leq \eta  \leq -1.6$ within $r\geq r_0$   \\
\hline
 $\rho(r) + P_r(r) + 2P_t(r)$ (Fig. \ref{fig6} right BP) &   $\rho(r) + P_r(r) + 2P_t(r) > 0$ for $-5/2 \leq \eta  \leq -3/2$ within $r\geq r_0$    \\
\hline
\end{tabular}
\end{table}

\begin{table}[thth]
\caption{ The behaviour of the energy conditions for the NFW DM model. }\label{tab2}
\centering
\begin{tabular}{|c|c|c|c|c|c|c|c|cc}
\hline
\multicolumn{2}{|c|}{For $\rho_s$ = 0.005, $ r_s = 2.1$, $ r_0 = 1$, and $\gamma$ = 2}\\
\hline
Expressions  & Results \\
\hline
 $\rho(r)$ (Fig. \ref{fig7} left above panel (AP)) & $\rho(r) > 0$ for $-3 \leq \lambda \leq 3$ within $r\geq r_0$ \\
\hline
 $\rho(r) + P_r(r)$ (Fig. \ref{fig7} middle AP) &  $\rho(r) + P_r(r) < 0$ for $-3 \leq \lambda<2$; $\rho(r) + P_r(r) \geq 0$ for $2 \leq \lambda \leq 3$  within $r\geq r_0$  \\
\hline
 $\rho(r) + P_t(r)$ (Fig. \ref{fig7} right AP) &  $\rho(r) + P_t(r) > 0$ for $-3 \leq \lambda \leq 3$ within $r\geq r_0$ \\
\hline
 $\rho(r) - P_r(r)$  (Fig. \ref{fig7} left below panel (BP)) &  $\rho(r) - P_r(r) > 0$ for $-3 \leq \lambda \leq 3$  within $r\geq r_0$   \\
\hline
 $\rho(r) - P_t(r)$ (Fig. \ref{fig7} middle BP) &  $\rho(r) - P_t(r) < 0$ for $-3 \leq \lambda \leq -0.24$; then (+)ve near $r=r_0$  \\
\hline
 $\rho(r) + P_r(r) + 2P_t(r)$ (Fig. \ref{fig7} right BP) &   $\rho(r) + P_r(r) + 2P_t(r) > 0$ for $-3 \leq \lambda \leq 3$  within $r\geq r_0$    \\
\hline
\hline
\multicolumn{2}{|c|}{For $\rho_s$ = 0.005, $ r_s = 2.1$, $ r_0 = 1$, and $\lambda$ = 1}\\
\hline
 $\rho(r)$ (Fig. \ref{fig8} left above panel (AP)) & $\rho(r) > 0$ for $1 \leq \gamma  \leq 2.5$ within $r\geq r_0$ \\
\hline
 $\rho(r) + P_r(r)$ (Fig. \ref{fig8} middle AP) &  $\rho(r) + P_r(r) < 0$ for $1 \leq \gamma  \leq 2.5$ within $r\geq r_0$  \\
\hline
 $\rho(r) + P_t(r)$ (Fig. \ref{fig8} right AP) &  $\rho(r) + P_t(r) > 0$ for $1 \leq \gamma  \leq 2.5$ within $r\geq r_0$ \\
\hline
 $\rho(r) - P_r(r)$  (Fig. \ref{fig8} left below panel (BP)) &  $\rho(r) - P_r(r) > 0$ for $1 \leq \gamma  \leq 2.5$ within $r\geq r_0$   \\
\hline
 $\rho(r) - P_t(r)$ (Fig. \ref{fig8} middle BP) &  $\rho(r) - P_t(r) < 0$ for $1 \leq \gamma  \leq 1.26$; then the maximum (+)ve region near $r=r_0$  \\
\hline
 $\rho(r) + P_r(r) + 2P_t(r)$ (Fig. \ref{fig8} right BP) &   $\rho(r) + P_r(r) + 2P_t(r) > 0$ for $1 \leq \gamma  \leq 2.5$ within $r\geq r_0$    \\
\hline
\end{tabular}
\end{table}

\subsection{  Active Gravitational Mass}\label{sec7c}
Here, we are interested in studying the active gravitational mass for the reported wormhole solutions supported by the King and NFW DM models.
The active gravitational mass function $M_A$ for the wormhole, within the region $r_0 \leq r \leq R$, is expressed as
\begin{eqnarray}
    M_A = 4\pi \int_{r_0}^r \bar{r}^2\rho(\bar{r})d\bar{r}.
\end{eqnarray}
It is worth noting that a positive active gravitational mass signifies the physical acceptability of the models. 

For the King DM density profile, we obtain the active gravitational mass in the following form
\begin{eqnarray}
    M_A = \left[\frac{4}{3} \pi  \alpha  r^3 \left(\frac{r^2}{r_s^2 \beta }+1\right)^{-\eta } \left(\frac{r^2}{r_s^2}+\beta \right)^{\eta } \, _2F_1\left(\frac{3}{2},-\eta ;\frac{5}{2};-\frac{r^2}{r_s^2 \beta }\right)\right]_{r_0}^r.
\end{eqnarray}
For the NFW DM density profile, the active gravitational mass is obtained as
\begin{eqnarray}
    M_A = \left[\frac{4\pi\rho _s}{3-\gamma } r^3\left(\frac{r}{r_s}\right)^{-\gamma} \, _2F_1\left(3-\gamma ,3-\gamma ;4-\gamma ;-\frac{r}{r_s}\right)\right]_{r_0}^r.
\end{eqnarray}
In the wormhole scenarios under consideration, both mass functions are positive and increase with radius, as evident from Fig. \ref{fig13}. Therefore, the proposed wormholes models are physically acceptable.

\begin{figure}[!htbp]
\begin{center}
\begin{tabular}{rl}
\includegraphics[width=6cm]{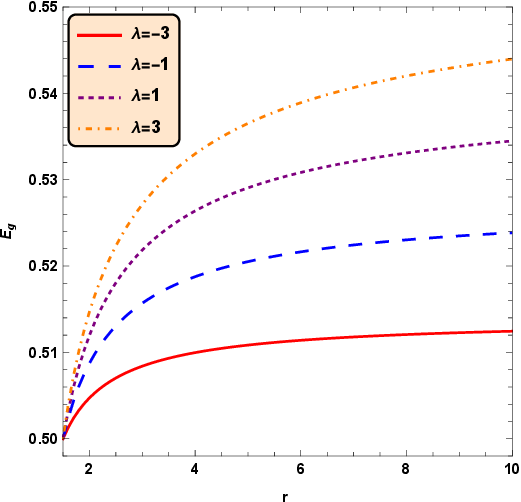}
\includegraphics[width=6cm]{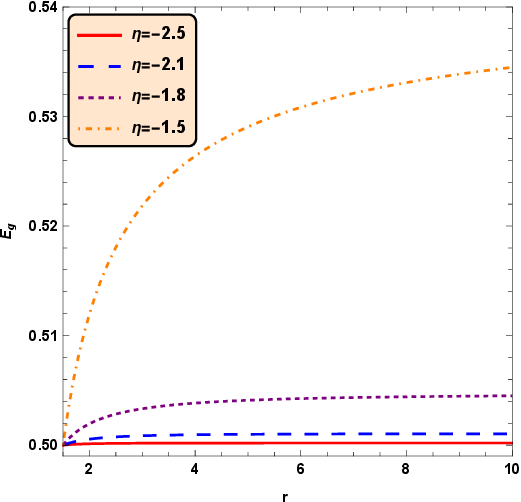}
\\
\end{tabular}
\begin{tabular}{rl}
\includegraphics[width=6cm]{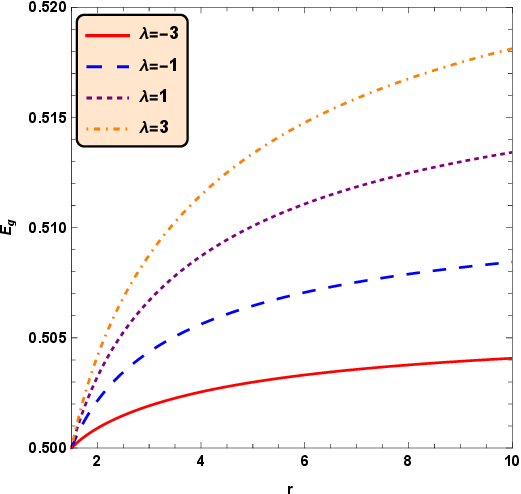}
\includegraphics[width=6cm]{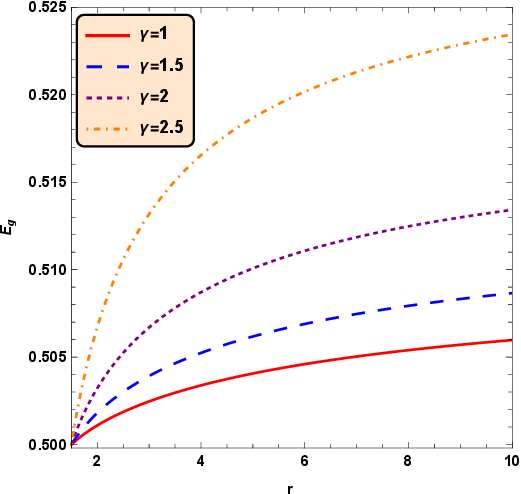}
\\
\end{tabular}
\end{center}
\caption{ Shows the radial variations of the total gravitational energy under the King DM model with $\alpha$ = 1.5, $ r_s = 0.3$, $ r_0 = 1$, $\beta$ = 1, and $\eta$ = -3/2 (Left), and $\alpha$ = 1.5, $ r_s = 0.3$, $ r_0 = 1$, $\beta$ = 1, and $\lambda$ = 1 (Right) in the above panel; under the NFW DM model with $\rho_s$ = 0.005, $ r_s = 2.1$, $ r_0 = 1$, and $\gamma$ = 2 (Left), and $\rho_s$ = 0.005, $ r_s = 2.1$, $ r_0 = 1$, and $\lambda$ = 1 (Right) in the below panel.}\label{fig14}
\end{figure}

\subsection{   Total Gravitational Energy}\label{sec7d}

The foregoing discussion has established that the matter content of our proposed wormholes violates the radial NEC in most of the regions of model parameters and is thus classified as exotic matter. Subsequently, the total gravitational energy of a structure composed of normal baryonic matter is negative. Therefore, we are willing to examine the nature of gravitational energy in the scenario of the proposed wormholes. The total gravitational energy $E_g$ of the present DM galactic wormholes can be expressed as \cite{dl07, kk09}
\begin{eqnarray}
  E_g = Mc^2 -E_M,\label{eg}  
\end{eqnarray}
where $Mc^2$ stands for the total energy, reads as
\begin{eqnarray}
    Mc^2 = \frac{1}{2}\int_{r_0}^r\left(T_t^t\right)^Mr^2dr+\frac{r_0}{2}.
\end{eqnarray}
In the above expression for $Mc^2$, the term $\frac{r_0}{2}$ is associated with the effective mass \cite{kk09}, and $E_M$ represents the total of other energy components, such as rest energy, internal energy, kinetic energy, and so on, and is given by
\begin{eqnarray}
    E_M = \frac{1}{2}\int_{r_0}^r r^2\left(T_t^t\right)^M \sqrt{g_{rr}} dr, ~~ \text{with}~~ g_{rr} = \left(1-\frac{\Omega(r)}{r}\right)^{-1}. 
\end{eqnarray}

Thus, the total gravitational energy (\ref{eg}) reads as
\begin{eqnarray}
  E_g = \frac{1}{2}\int_{r_0}^r\left[1-\sqrt{g_{rr}}\right]\left(T_t^t\right)^Mr^2dr+\frac{r_0}{2}.\label{eg1}  
\end{eqnarray}

 Misner \cite{cw73} proposed that the total gravitational energy is attractive if $E_g < 0$ and repulsive if $E_g > 0$.  It is quite difficult to obtain exact solutions of the integral (\ref{eg1}) due to its complicated form. Therefore, we have used the numerical technique to solve it and demonstrated graphically in Fig. \ref{fig14}. The results of  Fig. \ref{fig14} ensure that $E_g > 0$ for both King's and NFW models, which signifies the repulsive nature of total gravitational energy in the vicinity of the wormhole throat. Indeed, this repulsive nature of the total gravitational energy is consistent with the formation of physically viable galactic wormholes under the framework of Kalb-Ramond gravity. 

\begin{figure}[!htbp]
\begin{center}
\begin{tabular}{rl}
\includegraphics[width=5.8cm]{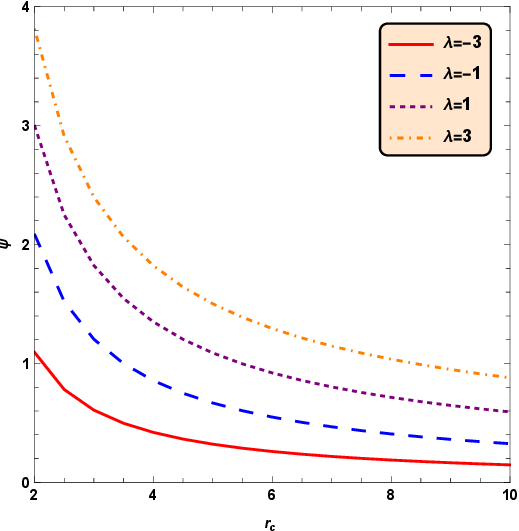}
\includegraphics[width=6cm]{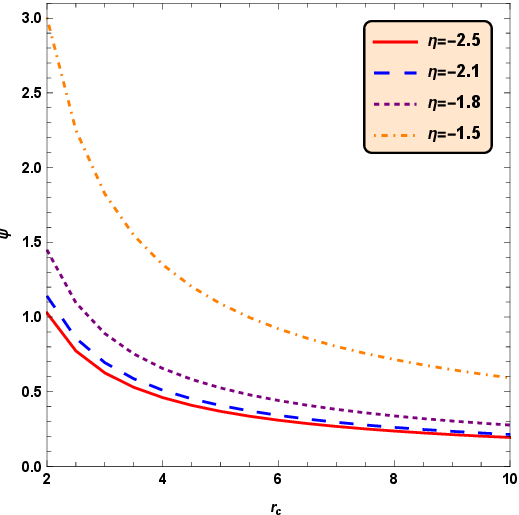}
\\
\end{tabular}
\begin{tabular}{rl}
\includegraphics[width=6cm]{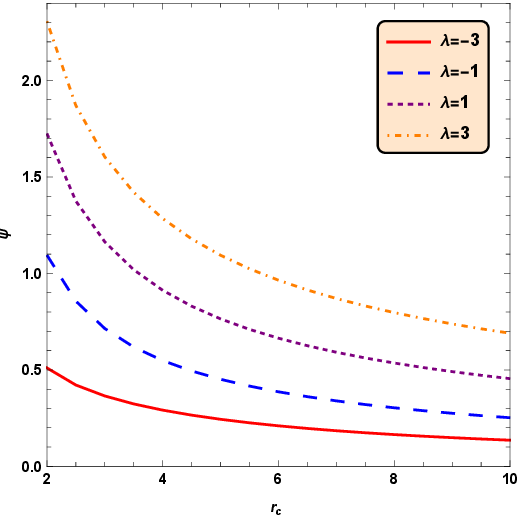}
\includegraphics[width=6cm]{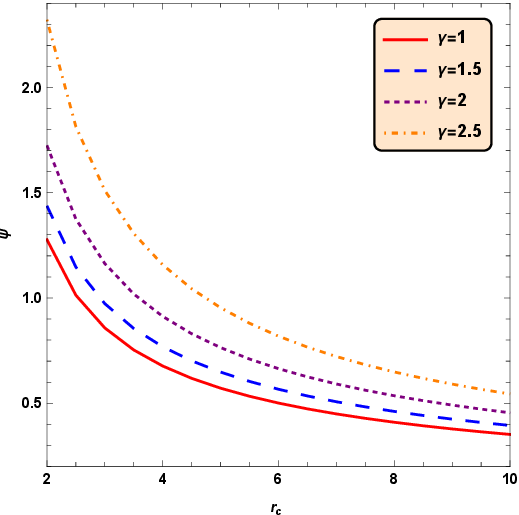}
\\
\end{tabular}
\end{center}
\caption{ Shows the variations of the deflection angels against $r_c$ under the King DM model with $\alpha$ = 1.5, $ r_s = 0.3$, $ r_0 = 1$, $\beta$ = 1, $\eta$ = -3/2, and $A = 0.1$ (Left), and $\alpha$ = 1.5, $ r_s = 0.3$, $ r_0 = 1$, $\beta$ = 1, $\lambda$ = 1, and $A = 0.1$ (Right) in the above panel; under the NFW DM model with $\rho_s$ = 0.005, $ r_s = 2.1$, $ r_0 = 1$, $\gamma$ = 2, and $A = 0.1$. (Left), and $\rho_s$ = 0.005, $ r_s = 2.1$, $ r_0 = 1$, $\lambda$ = 1, and $A = 0.1$ (Right) in the below panel.}\label{fig15}
\end{figure}

\section{DEFLECTION ANGLE}\label{sec9}

This section focuses on the investigation of deflection angle produced by the wormholes under the influence of the distinct DM density profiles considered. The literature establishes that mass and energy distort the fabric of space-time, creating curvature that governs the motion of objects and light. In the presence of space-time curvature, the path of light bends near massive objects like black holes or wormholes, even though its speed remains constant.  This phenomenon results in gravitational lensing, highlighting the profound influence of gravity on the trajectory of light. The interest in gravitational lensing, particularly in its strong form, has grown significantly among researchers, especially following the influential works of Virbhadra and collaborators \cite{ks98, ks00, cm01}. Their studies have paved the way for using strong gravitational lensing as a tool to probe the nature of space-time and explore astrophysical objects. Furthermore, Bozza \cite{vb02} developed an analytical framework for calculating gravitational lensing in the strong-field limit applicable to any spherically symmetric space-time. Notably, Bozza’s method has found extensive application in later works, such as those presented in Refs. \cite{jp03, jm12}. Motivated by this background, we employ Bozza’s analytical technique in the present study to evaluate the deflection angles produced by the proposed wormholes. The deflection angle $\psi$ for the Morris-Thorne wormhole structure is given by the following  expression \cite{vb01, vb02}
\begin{eqnarray}
\psi=-\pi+2\int_{r_c}^{\infty} \frac{e^{\Phi(r)}}{r^2\sqrt{\left(1-\frac{\Omega(r)}{r}\right)\left(\frac{1}{\beta^2}-\frac{e^{2\Phi(r)}}{r^2}\right)}} dr. \label{DA}
\end{eqnarray}
Here, $r_c$ denotes the closest approach of the light ray to the throat, and $\beta$ represents the impact parameter. For null geodesic, the relationship between $\beta$ and $r_c$ is given by
\begin{eqnarray}
\beta=r_c e^{-\Phi(r_c)}. \label{beta}
\end{eqnarray}

For our proposed wormhole solutions with constant redshift function $\Phi(r) = A = cont.$, and shape functions (\ref{B}) and (\ref{B1}), we compute the deflection angle from Eq. (\ref{DA}) as functions of the closest approach $r_c$. It is noted that we adopt the numerical technique to study the deflection angle in the vicinity of the wormhole's throat. The deflection angle for both King's and NFW models decreases as $r_c$ increases, while it increases with the increasing model parameters $\lambda$, $\eta$, and $\gamma$, clear from Fig. \ref{fig15}. In addition, the deflection angle approaches zero as the distance $r_c$ tends to infinity, i.e., far from the wormhole's throat, where the wormhole's gravity is negligible, the light ray experiences no significant deflection from its original path. Moreover, the deflection angle rises sharply in the vicinity of the wormhole throat and tends to infinity at the wormhole throat, where the gravitational field is extremely strong.

\section{Results and Conclusion}\label{sec10}
In this study, we have explored asymptotically flat traversable wormhole solutions supported by two distinct DM density models, the King model and the Navarro-Frenk-White (NFW) model, within the framework of Kalb-Ramond gravity. In particular, we have employed the DM density profile equations to derive the shape functions that characterise the wormhole geometries. The obtained shape functions are mostly monotonically increasing and remain less than the radial coordinate $r$ after the wormhole throat, i.e. $b(r)/r < 1$ for $r > r_{0}$ with $b(r)/r \rightarrow 0$ when $r\rightarrow \infty$. They also satisfy the flare-out condition, as illustrated in Figs. \ref{fig1}-\ref{fig4}. Consequently, the proposed shape functions yield valid wormhole geometries by fulfilling all the fundamental criteria required for the existence of a traversable wormhole. It is important to emphasize that the model parameters play a crucial role in determining the wormhole geometry. Our analysis reveals that appropriate parameter choices lead to asymptotically flat traversable wormhole solutions that satisfy all necessary conditions at the throat. Further, we have analyzed the energy conditions corresponding to the present models, employing the same set of parameters used in the construction of the shape functions. In particular, the mathematical expressions for the Null Energy Condition (NEC) at the throat have been derived for each model, as presented in Eq. (\ref{nec0}). The regions where the energy conditions are satisfied or violated are visually represented in Figs. \ref{fig5}–\ref{fig8} and comprehensively summarized in Tables- \ref{tab1} and \ref{tab2}. In the analysis of energy conditions, it has been found that the model parameters $\lambda$, $\eta$, and $\gamma$ play a pivotal role in determining their behaviour. Within the considered ranges of parameters, all the energy conditions are largely satisfied except NEC. For both wormhole models, the radial NEC is violated for $-3 \leq \lambda < 2$, implying the presence of exotic matter that contributes to the stability and traversability of wormholes by counterbalancing the gravitational collapse typically caused by ordinary matter. In contrast,  the radial NEC is satisfied for $2 \leq \lambda \leq 3$, indicating that nonexotic, physically acceptable matter can support the wormhole structure within this parameter range. Thus, the Kalb-Ramond gravity framework admits wormhole solutions supported by both exotic and nonexotic matter, depending on the values of the Kalb-Ramond gravity parameter $\lambda$. This dual capability underscores the versatility of Kalb-Ramond gravity in accommodating a wide spectrum of matter content while maintaining the conditions required for traversable wormholes. It is important to note that wormhole solutions in standard Einstein gravity typically require the presence of exotic matter to sustain the geometry \cite{ks88}. However, various studies have shown that within the framework of modified gravity theories, the existence of exotic matter is not always necessary for constructing wormhole structures \cite{mk15, gc12, mf20, ns25}. In addition, the proposed wormhole configurations maintain their equilibrium scenario under the combined influence of hydrostatic and anisotropic forces, as illustrated in Fig. \ref{fig9}. Moreover, we have investigated several physical key features of the present wormholes, such as the embedding surface, complexity factor, active gravitational mass, and total gravitational energy. The embedding surfaces $z(r)$ for proposed wormholes are demonstrated in Fig. \ref{fig10} for King's model and in Fig. \ref{fig11} for NFW model, along with the complete visual representation of the wormhole structure. In these diagrams, the positive curvature  ($z > 0$) corresponds to the upper universe, while negative curvature ($z < 0$) represents the lower universe of the wormholes. The complexity factor for the present wormhole models approaches zero at large radial distances from the wormhole throat, as shown in Fig. \ref{fig12},  indicating the wormhole configurations with inhomogeneous energy density and anisotropic pressure. The active gravitational mass functions for both wormhole models are positive and increase monotonically with the radial coordinate $r$, confirming the physical viability of the solutions (see Fig. \ref{fig13}). Furthermore, the total gravitational energy is found to be repulsive from the graphical demonstration in Fig. \ref{fig14}, supporting the plausibility of forming physically realistic galactic DM supporting wormholes in the Kalb-Ramond gravity framework. 

We have also carried out a comprehensive analysis of gravitational lensing, with particular focus on the strong lensing effects induced by the wormhole geometry in the context of each considered DM model. To this end, we have employed Bozza’s formalism for gravitational lensing in the strong-field regime, applicable to a general spherically symmetric metric \cite{vb02}. Our findings reveal that the deflection angle for both the King and NFW dark matter models exhibits a decreasing trend with increasing closest approach distance $r_c$, and the increasing trend with increasing values of the model parameters $\lambda$, $\eta$, and $\gamma$ (see Fig. \ref{fig15}). This behaviour suggests that the larger impact parameter as well as the smaller model parameters reduce the influence of the wormhole’s gravitational field on the trajectory of light. Additionally, the deflection angle asymptotically approaches zero as $r_c\rightarrow \infty$, indicating that at large distances from the wormhole throat, where the gravitational field becomes negligibly weak, light rays propagate essentially along straight paths, unaffected by the wormhole geometry. Conversely, as the light ray approaches the wormhole throat, the deflection angle increases rapidly, ultimately diverging at the throat. This divergence signifies a region where the gravitational field is extremely strong and capable of trapping light in unstable circular orbits. Indeed, the sharp rise in deflection angle near the throat not only highlights the intense curvature of spacetime in that region but also underscores the observational relevance of strong lensing signatures in detecting and characterising traversable wormholes.
  
In recent years, observational evidence has increasingly supported the existence of black holes at the centers of galaxies. However, no direct observational evidence for the existence of wormholes has been found to date. As a result, the scientific community continues to actively search for potential observational signatures that can confirm their existence. In this context, over the past few decades, a wide range of theoretical models for wormholes have been proposed, not only within the framework of Einstein's general relativity but also in various modified gravity theories and even at galactic scales. In particular, the galactic wormhole solutions have been derived based on several DM density profiles, such as the NFW DM density profile \cite{fr16}, URC DM density profile\cite{fr14}, Bose-Einstein DM density profile \cite{kj19}, King's DM density profile \cite{si19}. Moreover, the DM supporting galactic wormholes have also been investigated in the framework of $f(T)$ gravity \cite{sm14} and $f(G)$ gravity \cite{sm16}, teleparallel gravity \cite{gm24},  Einstein cubic gravity \cite{gm23}, and  4D Einstein-Gauss-Bonnet gravity \cite{zh24}. These findings collectively suggest that both the King and NFW DM density profiles are well-suited to support galactic wormhole configurations. Besides, the study of light deflection by wormholes has attracted considerable attention from researchers in recent years \cite{wj22, wj22a, yk21, ao20, wj19, ao18, kj18}.  In this study, we have employed the King and NFW DM models to construct new wormhole structures within the framework of  Kalb-Ramond gravity. Interestingly, the resulting wormhole solutions satisfy all the necessary geometric and physical conditions,  including the flare-out criterion, energy conditions depending on $\lambda$, and equilibrium under internal forces, thereby reinforcing their viability within the galactic DM halo under the framework of  Kalb-Ramond gravity. Furthermore, we have investigated the deflection angles associated with these wormholes for each DM model, representing a novel contribution within the context of modified theories of gravity. It is important to note that the present wormhole solutions are purely theoretical. As a result, the scientific community continues to seek observational evidence for the existence of such wormholes, particularly through the study of scalar wave scattering, gravitational lensing \cite{pk14}, and the analysis of gamma-ray burst light curves \cite{dt98}.

Finally, based on the comprehensive analysis carried out in this study, it is reasonable to conclude that our findings offer strong theoretical support for the possible existence of the King and NFW DM supporting traversable wormholes with the presence of exotic or nonexotic matter under the framework of Kalb-Ramond gravity. Furthermore, these DM profiles may serve as viable inputs for constructing wormhole solutions in other modified theories of gravity, potentially broadening the scope of their applicability in alternative gravitational frameworks.

\end{document}